%% file: gershon-SUSSP.tex
\documentclass[graybox]{svmult}

\usepackage{mathptmx}       
\usepackage{helvet}         
\usepackage{courier}        
\usepackage{type1cm}        
\usepackage{makeidx}         
\usepackage{graphicx}        
\usepackage{multicol}        
\usepackage[bottom]{footmisc}

\usepackage{xspace} 
\usepackage{ifthen} 
\newboolean{articletitles}
\setboolean{articletitles}{false} 
\usepackage{upgreek} 
\newboolean{uprightparticles}
\setboolean{uprightparticles}{false} 
\input{lhcb-symbols-def} 

\usepackage{hyperref}    
\usepackage[all]{hypcap} 

\usepackage{subdepth}

\usepackage{cite} 
\usepackage{mciteplus}

\makeindex             

\begin{document}

\title*{Flavour Physics in the LHC Era}
\author{Tim Gershon}
\institute{
  Tim Gershon \at
  Department of Physics, University of Warwick, Coventry, United Kingdom
  {\it and}
  European Organization for Nuclear Research (CERN), Geneva, Switzerland, 
  \email{T.J.Gershon@warwick.ac.uk}
}

\maketitle

\abstract{
  These lectures give a topical review of heavy flavour physics, in particular \CP violation and rare decays, from an experimental point of view.
  They describe the ongoing motivation to study heavy flavour physics in the LHC era, the current status of the field emphasising key results from previous experiments, some selected topics in which new results are expected in the near future, and a brief look at future projects.
}

\section{Introduction}
\label{sec:intro}

%
The concept of ``flavour physics'' was introduced in the 1970s~\cite{Browder:2008em}
\begin{quotation}
  The term flavor was first used in particle physics in the context of the quark model of hadrons. It was coined in 1971 by Murray Gell-Mann and his student at the time, Harald Fritzsch, at a Baskin-Robbins ice-cream store in Pasadena. Just as ice cream has both color and flavor so do quarks.
\end{quotation}
Leptons also come in different flavours, and flavour physics covers the properties of both sets of fermions.
Counting the fundamental parameters of the Standard Model (SM), the 3 lepton masses, 6 quark masses and 4 quark mixing (CKM) matrix~\cite{Cabibbo:1963yz,Kobayashi:1973fv} parameters are related to flavour physics.  
In case neutrino masses are introduced, the new parameters (at least 3 more masses and 4 more mixing parameters) are also related to flavour physics.
This large number of free parameters is behind several of the mysteries of the SM:
\begin{itemize}
\item Why are there so many different fermions?
\item What is responsible for their organisation into generations / families?
\item Why are there 3 generations / families each of quarks and leptons?
\item Why are there flavour symmetries?
\item What breaks the flavour symmetries?
\item What causes matter -- antimatter asymmetry?
\end{itemize}

Unfortunately these mysteries will not be answered in these lectures -- they are mentioned here simply because it is important to bear in mind their existence.
Instead the focus will be on specific topics in the flavour-changing interactions of the charm and beauty quarks,\footnote{
  It is one of the peculiarities of our field that ``heavy flavour physics'' does not include discussion of the heaviest flavoured particle, the top quark.
} with occasional digressions on related topics.

While our main interest is in the properties of the charm and beauty quarks, due to the strong interaction, experimental studies must be performed using one or more of the many different charmed or beautiful hadrons.
These can decay to an even larger multitude of different final states, making learning the names of all the hadrons a big challenge for flavour physicists.
Moreover, hadronic effects can often obscure the underlying dynamics.
Nevertheless, it is the hadronisation that results in the very rich phenomenology that will be discussed, so one should bear in mind that~\cite{Bigi:2005av}
\begin{quotation}
  The strong interaction can be seen either as the ``unsung hero'' or the ``villain'' in the story of quark flavour physics.
\end{quotation}

\section{Motivation to study heavy flavour physics in the LHC era}
\label{sec:motivation}
 
There are two main motivations for ongoing experimental investigations into heavy flavour physics: (i) \CP violation and its connection to the matter-antimatter asymmetry of the Universe; (ii) discovery potential far beyond the energy frontier via searches for rare or SM forbidden processes.
These will be discussed in turn below.

First let us consider one of the mysteries listed above (What breaks the flavour symmetries?) to see how it is connected to these motivations.
In the SM, the vacuum expectation value of the Higgs field breaks the electroweak symmetry.
Fermion masses arise from the Yukawa couplings of the quarks and charged leptons to the Higgs field, and the CKM matrix arises from the relative misalignment of the Yukawa matrices for the up- and down-type quarks.
Consequently, the only flavour-changing interactions are the charged current weak interactions.
This means that there are no flavour-changing neutral currents (the GIM mechanism~\cite{Glashow:1970gm}), a feature of the SM which is not generically true in most extended theories.
Flavour-changing processes provide sensitive tests of this prediction; as an example, many new physics (NP) models induce contributions to the $\mu \to e\gamma$ transition at levels close to (or even above!) the current experimental limit, recently made more restrictive by the MEG experiment~\cite{Adam:2013mnn},
${\cal B}(\mu^+ \to e^+\gamma) < 5.7 \times 10^{-13}$ at 90\% confidence level (CL).
Improved experimental reach in this and related charged lepton flavour violation searches therefore provides interesting and unique NP discovery potential (for a review, see, \eg, Ref.~\cite{Marciano:2008zz}).

\subsection{\CP violation}
\label{subsec:CPV}

As mentioned above, the CKM matrix arises from the relative misalignment of the Yukawa matrices for the up- and down-type quarks:
\begin{equation}
  V_{CKM} = U_u U^\dagger_d \, ,
\end{equation}
where $U_u$ and $U_d$ diagonalise the up- and down-type quark mass matrices respectively.
Hence, $V_{CKM}$ is a $3\times3$ complex unitary matrix.
Such a matrix is in general described by 9 (real) parameters, but 5 can be absorbed as unobservable phase differences between the quark fields.
This leaves 4 parameters, of which 3 can be expressed as Euler mixing angles, but the fourth makes the CKM matrix complex -- and hence the weak interaction couplings differ for quarks and antiquarks, \ie\ \CP violation arises.

The expression ``\CP violation'' refers to the violation of the symmetry of the combined $C$ and $P$ operators, which replace particle with antiparticle (charge conjugation) and invert all spatial co-ordinates (parity) respectively.  
Therefore \CP violation provides absolute discrimination between particle and antiparticle: one cannot simply swap the definition of which is called ``particle'' with a simultaneous redefinition of left and right.\footnote{
  The importance of \CP violation in this regard was noted by Landau~\cite{Landau:1957tp} following the observation of parity violation~\cite{Lee:1956qn,Wu:1957my}.
}
There is a third discrete symmetry, time reversal ($T$), and it is important to note that there is a theorem that states that \CPT must be conserved in any locally Lorentz invariant quantum field theory~\cite{Lueders:1992dq}.  
Therefore, under rather reasonable assumptions, an observation of \CP violation corresponds to an observation of $T$ violation, and vice versa.
Nonetheless, it remains of interest to establish $T$ violation without assumptions regarding other symmetries~\cite{Angelopoulos:1998dv,Lees:2012uka}. 

The four parameters of the CKM matrix can be expressed in many different ways, but two popular choices are the Chau-Keung (PDG) parametrisation -- $(\theta_{12}, \theta_{13}, \theta_{23}, \delta)$~\cite{Chau:1984fp} -- and the Wolfenstein parametrisation -- $(\lambda, A, \rho, \eta)$~\cite{Wolfenstein:1983yz}.
In both cases a single parameter ($\delta$ or $\eta$) is responsible for all \CP violation.
This encapsulates the predictivity that makes the CKM theory such a remarkable success: it describes a vast range of phenomena at many different energy scales, from nuclear beta transitions to single top quark production, all by only four parameters (plus hadronic effects).

Let us digress a little into history.
In 1964, \CP violation was discovered in the kaon system~\cite{Christenson:1964fg}, but it was not until 1973 that Kobayashi and Maskawa proposed that the effect originated from the existence of three quark families~\cite{Kobayashi:1973fv}.
On a shorter time-scale, in 1967 Sakharov noted that \CP violation was one of three conditions necessary for the evolution of a matter-dominated universe, from a symmetric initial state~\cite{Sakharov:1967dj}:
\begin{enumerate}
\item baryon number violation,
\item $C$ and \CP violation,
\item thermal inequilibrium.
\end{enumerate}
This observation evokes the prescient concluding words of Dirac's 1933 Nobel lecture, discussing his successful prediction of the existence of antimatter, in the form of the positron~\cite{Dirac}:
\begin{quotation}
  If we accept the view of complete symmetry between positive and negative electric charge so far as concerns the fundamental laws of Nature, we must regard it rather as an accident that the Earth (and presumably the whole solar system), contains a preponderance of negative electrons and positive protons. 
  It is quite possible that for some of the stars it is the other way about, these stars being built up mainly of positrons and negative protons. 
  In fact, there may be half the stars of each kind. 
  The two kinds of stars would both show exactly the same spectra, and there would be no way of distinguishing them by present astronomical methods.
\end{quotation}
Dirac was not aware of the existence of \CP violation, that breaks the {\it complete symmetry} of the laws of Nature.
Moreover, modern astronomical methods do allow to search for antimatter dominated regions of the Universe, and none have been observed (though searches, for example by the PAMELA and AMS experiments, are ongoing).
Therefore, \CP violation appears to play a crucial role in the early Universe.

We can illustrate this with a simple exercise.  
Suppose we start with equal amounts of matter ($X$) and antimatter ($\bar{X}$).
The matter $X$ decays to final state $A$ (with baryon number $N_A$) with probability $p$ and to final state $B$ (baryon number $N_B$) with probability $(1-p)$.
The antimatter, $\bar{X}$, decays to final state $\bar{A}$ (with baryon number $-N_A$) with probability $\bar{p}$ and final state $\bar{B}$ (baryon number $-N_B$) with probability $(1-\bar{p})$.
The resulting baryon asymmetry is 
$$ \Delta N_{\rm tot} = N_A p + N_B(1-p) - N_A\bar{p} - N_B(1-\bar{p}) = (p - \bar{p}) (N_A - N_B) \, .$$
So clearly $ \Delta N_{\rm tot} \neq 0 $ requires both $p \neq \bar{p}$ and $N_A \neq N_B$, \ie\ both \CP violation and baryon number violation.

It is natural to next ask whether the magnitude of the baryon asymmetry of the Universe could be caused by the \CP violation in the CKM matrix.
The baryon asymmetry can be quantified relative to the number of photons in the Universe,
$$
\Delta N_B/N_\gamma = (N({\rm baryon}) - N({\rm antibaryon}))/N_\gamma \sim 10^{-10} \, .
$$
This can be compared to a dimensionless and parametrisation invariant measure of the amount of \CP violation in the SM, $J \times P_u \times P_d / M^{12}$,
where 
\begin{itemize}
\item $J = \cos(\theta_{12}) \cos(\theta_{23}) \cos^2(\theta_{13}) \sin(\theta_{12}) \sin(\theta_{23}) \sin (\theta_{13}) \sin(\delta)$ ,
\item $P_u = (m_t^2-m_c^2)(m_t^2-m_u^2)(m_c^2-m_u^2)$,
\item $P_d = (m_b^2-m_s^2)(m_b^2-m_d^2)(m_s^2-m_d^2)$,
\item and $M$ is the relevant scale, which can be taken to be the electroweak scale, ${\cal O}(100 \gev)$.
\end{itemize}
The parameter $J$ is known as the Jarlskog parameter~\cite{Jarlskog:1985ht}, and is expressed above in terms of the Chau-Keung parameters.
Putting all the numbers in, we find a value for the asymmetry of $\sim 10^{-17}$, much below the observed $10^{-10}$.
This is the origin of the widely accepted statement that the SM \CP violation is insufficient to explain the observed baryon asymmetry of the Universe.
Note that this occurs primarily not because $J$ is small, but rather because the electroweak mass scale is far above the mass of most of the quarks.  
Therefore, to explain the baryon asymmetry of the Universe, there must be additional sources of \CP violation that occur at high energy scales.
There is, however, no guarantee that these are connected to the \CP violation that we know about.
The new sources may show up in the quark sector via discrepancies with CKM predictions (as will be discussed below), but could equally appear in the lepton sector as \CP violation in neutrino oscillations.
Or, for that matter, new sources could be flavour-conserving and be found in measurements of electric dipole moments, or could be connected to the Higgs sector, or the gauge sector, or to extra dimensions, or to other NP.
In any case, precision measurements of flavour observables are generically sensitive to additions to the SM, and hence are well-motivated.

In this context, it is worth noting the enticing possibility of ``leptogenesis'', where the baryon asymmetry is created via a lepton asymmetry (see, \eg, Ref.~\cite{Davidson:2008bu} for a review). 
In the case that neutrinos are Majorana particles -- \ie\ they are their own antiparticles -- the right-handed neutrinos may be very massive, which provides an immediate connection with the needed high energy scale.
Experimental investigation of this concept requires the determination of the lepton mixing (PMNS)~\cite{Pontecorvo:1957cp,Maki:1962mu} matrix, and proof whether or not neutrinos are Majorana particles.
The recent determination of the neutrino mixing angle $\theta_{13}$~\cite{An:2012eh,Ahn:2012nd} provides an important step forward; the next challenges are to establish \CP violation in neutrino oscillations and to observe (or limit) neutrinoless double beta decay processes.

\subsection{Rare processes}
\label{subsec:RD}

We have already digressed into history, and we should avoid doing so too much, but it is striking how often NP has shown up at the precision frontier before ``direct'' discoveries at the energy frontier.
Examples include: the GIM mechanism being established before the discovery of charm; \CP violation being discovered and the CKM theory developed before the discovery of the bottom and top quarks; the observation of weak neutral currents before the discovery of the $Z$ boson.
In particular, loop processes are highly sensitive to potential NP contributions, since SM contributions are suppressed or absent.

\begin{figure}[!htb]
\centering
\includegraphics[width=0.4\textwidth]{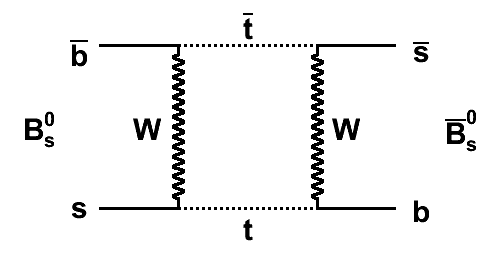}
\includegraphics[width=0.4\textwidth]{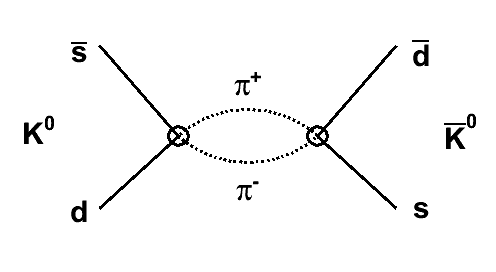}
\caption{
  Illustrative diagrams of (left) short-distance (dispersive) processes in \Bs mixing; (right) long-distance (absorptive) processes in $K^0$ mixing.
}
\label{fig:mix}
\end{figure}

As a specific example of this we can consider the loop processes involved in oscillations of neutral flavoured mesons.
(Rare decay processes will be discussed in more detail below.)
There are four such pseudoscalar particles in nature ($K^0$, $D^0$, $B^0$ and $B_s^0$) which can oscillate into their antiparticles via both short-distance (dispersive) and long-distance (absorptive) processes, as illustrated in Fig.~\ref{fig:mix}.
Representing such a meson generically by $M^0$, the evolution of the particle-antiparticle system is given by the time-dependent Schr\"odinger equation,
\begin{equation}
  i \frac{\partial}{\partial t} \left( \begin{array}{c} M^0 \\ \bar{M}^0 \end{array} \right) = \left( M - \frac{i}{2} \Gamma \right) \left( \begin{array}{c} M^0 \\ \bar{M}^0 \end{array} \right) \, ,
\end{equation}
where the effective\footnote{
  The complete Hamiltonian would include all possible final states of decays of $M^0$ and $\bar{M}^0$.
} Hamiltonian $H = M - \frac{i}{2} \Gamma$ is written in terms of $2\times2$ Hermitian matrices $M$ and $\Gamma$.
Note that the \CPT theorem requires that $M_{11} = M_{22}$ and $\Gamma_{11} = \Gamma_{22}$, \ie\ that particle and antiparticle have identical masses and lifetimes.

The physical states are eigenstates of the effective Hamiltonian, and are written
\begin{equation}
  \label{eq:pqdef}
  M_{\rm L,H} = p M^0 \pm q \bar{M}^0 \, ,
\end{equation}
where $p$ and $q$ are complex coefficients that satisfy $\left| p \right|^2 + \left| q \right|^2 = 1$.
Here the subscript labels L and H distinguish the eigenstates by their nature of being lighter or heavier; in some systems the labels S and L are instead used for short-lived and long-lived respectively (the choice depends on the values of the mass and width differences; the labels 1 and 2 are also sometimes used, usually to denote the \CP eigenstates).
\CP is conserved (in mixing) if the physical states correspond to the \CP eigenstates, \ie\ if $\left| q/p \right| = 1$.
Solving the Schr\"odinger equation gives
\begin{equation}
  \left( \frac{q}{p} \right)^2 = 
  \frac{M_{12}^* - \frac{i}{2}\Gamma_{12}^*}{M_{12} - \frac{i}{2}\Gamma_{12}} \, ,
\end{equation}
with eigenvalues given by $\lambda_{\rm L,H} = m_{\rm L,H} - \frac{i}{2}\Gamma_{\rm L,H} = (M_{11} - \frac{i}{2}\Gamma_{11}) \pm (q/p)(M_{12} - \frac{i}{2}\Gamma_{12})$, corresponding to mass and width differences $\Delta m = m_{\rm H} - m_{\rm L}$ and $\Delta \Gamma = \Gamma_{\rm H} - \Gamma_{\rm L}$ given by
\begin{eqnarray}
  (\Delta m)^2 - \frac{1}{4}(\Delta \Gamma)^2 & = & 4(\left|M_{12}\right|^2 + \frac{1}{4}\left|\Gamma_{12}\right|^2) \, ,\\
  \Delta m \Delta \Gamma & = & 4 \, {\rm Re}(M_{12}\Gamma_{12}^*) \, .
\end{eqnarray}
Note that with this notation, which is the same as that of Ref.~\cite{CPVreview}, $\Delta m$ is positive by definition while $\Delta \Gamma$ can have either sign.\footnote{
  With the definition given, $\Delta \Gamma$ is predicted to be negative for \Bd and \Bs mesons in the SM, and hence the sign-flipped definition is often encountered in the literature, \eg\ in Ref.~\cite{Bmixreview}.
}

Rather than going into the details of the formalism (which can be found in, \eg, Ref.~\cite{Nierste:2009wg}) let us instead take a simplistic picture.
\begin{itemize}
\item The value of $\Delta m$ depends on the rate of the mixing diagram of Fig.~\ref{fig:mix}(left).
  This depends on CKM matrix elements, together with various other factors that are either known or (in the case of decay constants and bag parameters) can be calculated using lattice QCD.  
  Moreover for the $B$ mesons, these other factors can be made to cancel in the $\Delta m_d/\Delta m_s$ ratio, such that the measured value of this quantity gives a theoretically clean determination of $\left| V_{td} / V_{ts} \right|^2$.

\item The value of $\Delta \Gamma$, on the other hand, depends on the widths of decays of the meson and antimeson into common final states (such as \CP-eigenstates).
  Therefore, $\Delta \Gamma$ is large for the $K^0$ system, where the two pion decay dominates, small for $D^0$ and \Bd mesons, where the most favoured decays are to flavour-specific or quasi-flavour-specific final states, and intermediate in the \Bs system.
  
\item Finally \CP violation in mixing tends to zero (\ie\ $q/p \approx 1$) if ${\rm arg}(\Gamma_{12}/M_{12}) = 0$, $M_{12} \ll \Gamma_{12}$ or $M_{12} \gg \Gamma_{12}$. 
\end{itemize}

This simplistic picture is sufficient to explain qualitatively the experimental values of the mixing parameters given in Table~\ref{tab:mix}.
It should be noted that $\Delta \Gamma (\Bs)$ has become well-measured only very recently (as discussed below), and that the experimental sensitivity for the \CP violation parameters in all of the $D^0$, $\Bd$ and $\Bs$ systems is still far from that of the SM prediction, making improved measurements very well motivated.

\begin{table}
  \caption{
    Qualitative expectations and measured values for the neutral meson mixing parameters.
    Experimental results are taken from Refs.~\cite{Beringer:1900zz,KLreview,Amhis:2012bh}.
    The definition of $a_{\rm sl}$ is given in footnote~\ref{foot-asl}.
  }
  \label{tab:mix}
  \centering
\begin{tabular}{c@{\hspace{5mm}}c@{\hspace{5mm}}c@{\hspace{5mm}}c}
\hline\noalign{\smallskip}
& $\Delta m$  & $\Delta \Gamma$ & $\left| q/p \right|$ \\
& ({\small $x = \Delta m/\Gamma$}) & ({\small $y = \Delta \Gamma / (2\Gamma))$}) & ({\small $a_{\rm sl} \approx 1-\left|q/p\right|^2$}) \\
\noalign{\smallskip}\svhline\noalign{\smallskip}
$K^0$ & large & $\sim$ maximal & small \\
      & $\sim 500$ & $\sim 1$ & $(3.32 \pm 0.06) \times 10^{-3}$ \\
$D^0$ & small & small & small \\
      & $(0.63 \pm 0.19)\%$ & $(0.75 \pm 0.12)\%$ & $0.52 \,^{+0.19}_{-0.24}$ \\
$\Bd$ & medium & small & small \\
      & $0.770 \pm 0.008$ & $0.008 \pm 0.009$ & $-0.0003 \pm 0.0021$ \\
$\Bs$ & large & medium & small \\
      & $26.49 \pm 0.29$ & $0.075 \pm 0.010$ & $-0.0109 \pm 0.0040$ \\ 
\noalign{\smallskip}\hline\noalign{\smallskip}
\end{tabular}
\end{table}

\begin{figure}[!htb]
\centering
\includegraphics[width=0.4\textwidth]{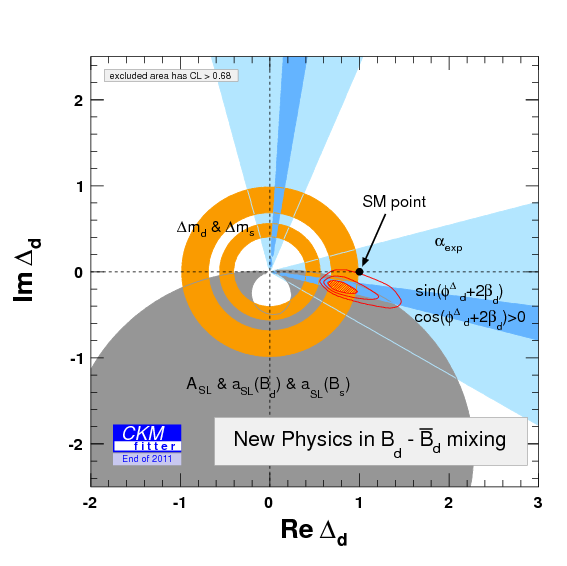}
\includegraphics[width=0.4\textwidth]{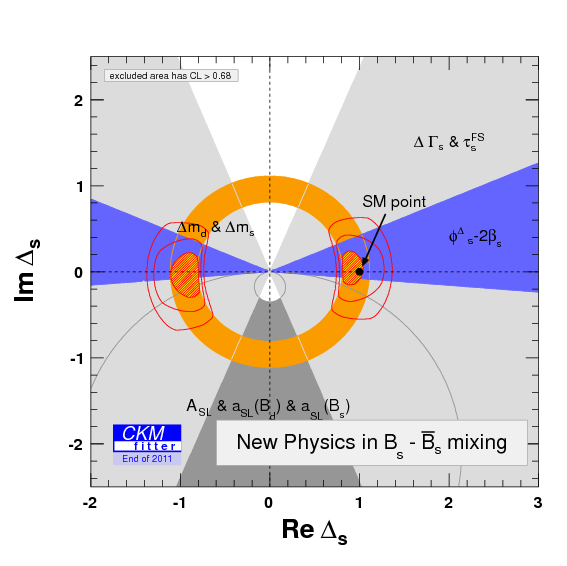}
\caption{
  Constraints on NP contributions in (left) \Bd and (right) \Bs mixing~\cite{Lenz:2012az}.
}
\label{fig:ckmfitNP}
\end{figure}

Thus, neutral meson oscillations are rare processes described by parameters that can be both predicted in the SM and measured experimentally.
All measurements to date are consistent with the SM predictions (though see below).
These results can then be used to put limits on non-SM contributions.
This can be done within particular models, but the model-independent approach, described in, \eg, Ref.~\cite{Isidori:2010kg} is illustrative.
The NP contribution is expressed as a perturbation to the SM Lagrangian,
\begin{equation}
  \label{eq:Leff}
  {\cal L}_{\rm eff} = {\cal L}_{\rm SM} + \Sigma \frac{c_i^{(d)}}{\Lambda^{d-4}}{\cal O}_{i}^{d}({\rm SM \ fields}) \, ,
\end{equation}
where the dimension $d$ of higher than 4 has an associated scale $\Lambda$ and couplings $c_i$.\footnote{
  In Eq.~(\ref{eq:Leff}) it is assumed that the NP modifies the SM operators; more generally extensions to the operator basis are also possible.
}
Given the observables in a given neutral meson system, NP contributions described effectively as four-quark operators ($d = 6$) can be constrained, either by putting bounds on $\Lambda$ for a fixed value of $c_i$ (typically 1), or by putting bounds on $c_i$ for a fixed value of $\Lambda$ (typically $1 \tev$).
In the former case bounds of ${\cal O}(100 \tev)$ are obtained; in the latter case the bounds can be ${\cal O}(10^{-9})$ or below~\cite{Isidori:2010kg}, with the strongest (weakest) bounds being in the $K^0$ ($\Bs$) sectors.
A similar analysis, but with more up-to-date inputs has been performed in Ref.~\cite{Lenz:2012az}, with results illustrated in Fig.~\ref{fig:ckmfitNP}.
The mixing amplitude, normalised to its SM value, is denoted by $\Delta$, and experimental constraints give $({\rm Re}\Delta, {\rm Im}\Delta)$ consistent with $(1,0)$ (\ie\ with the SM) for both \Bd and \Bs systems.

This is a very puzzling situation.
Limits on the NP scale give values of at least $100 \tev$ for generic couplings.
But, as discussed elsewhere, we expect NP to appear at the \tev scale to solve the hierarchy problem (and to provide a dark matter candidate, \etc)
If NP is indeed at this scale, NP flavour-changing couplings must be small.
But why?
This is the so-called ``new physics flavour problem''.

A theoretically attractive solution to this problem, known as minimal flavour violation (MFV)~\cite{D'Ambrosio:2002ex}, exploits the fact that the SM flavour-changing couplings are also small.
Therefore, if there is a perfect alignment of the flavour violation in a NP model with that in the SM, the tension is reduced.
The MFV paradigm is highly predictive, stating that there are no new sources of \CP violation and also that the correlations between certain observables share their SM pattern (the ratio of branching fractions of $\Bd \to \mumu$ and $\Bs \to \mumu$ being a good example).
Therefore, once physics beyond the SM is discovered, it will be an important goal to establish whether or not it is minimally flavour violating.
This further underlines that the flavour observables carry information about physics at very high scales.
 
\begin{figure}[!htb]
\centering
\includegraphics[width=0.4\textwidth]{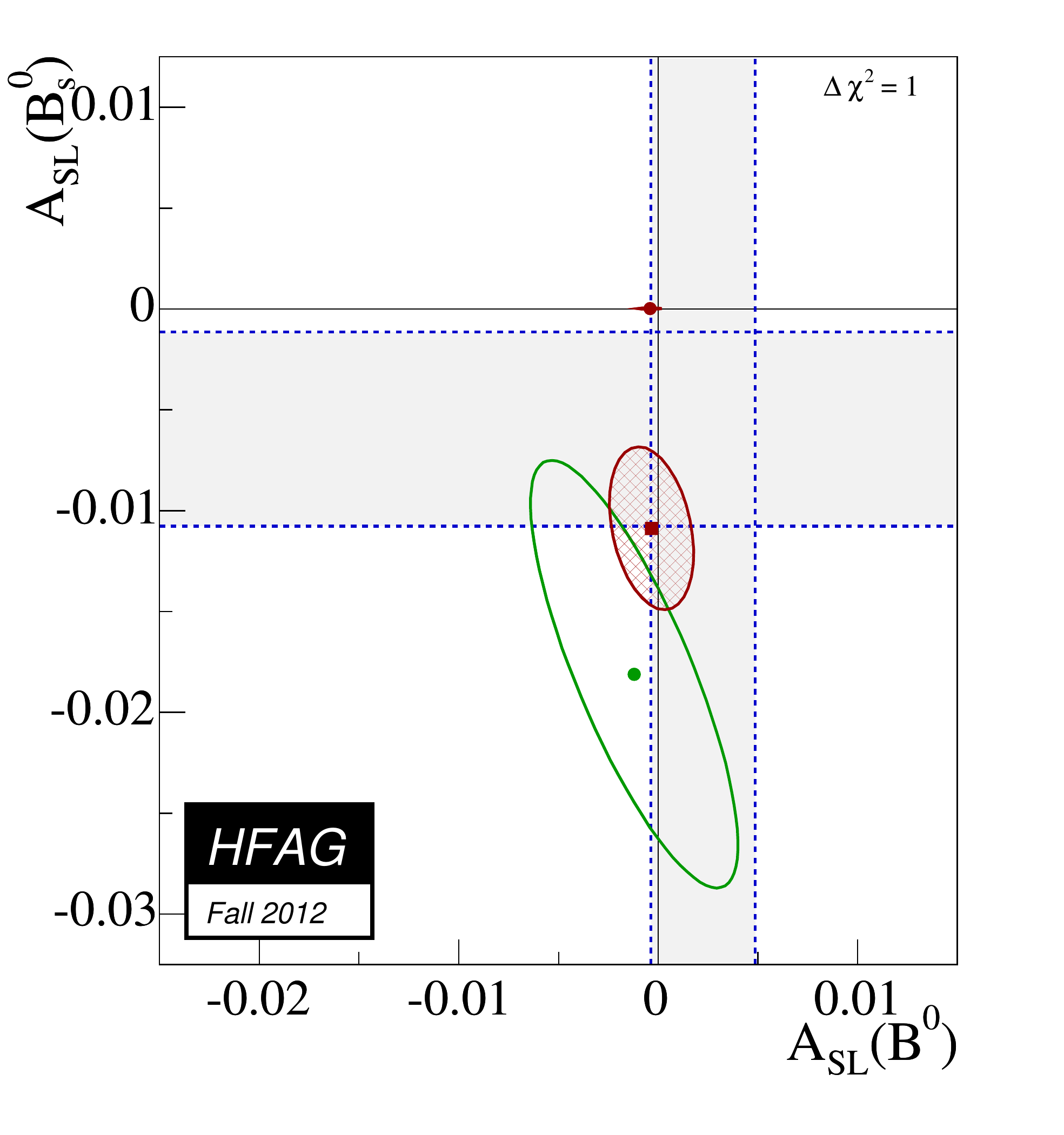}
\caption{
  World average of constraints on the parameters describing \CP violation in \Bd and \Bs mixing, $a_{\rm sl}^d$ and $a_{\rm sl}^s$.
  The green ellipse comes from the D0 inclusive same-sign dimuon analysis~\cite{Abazov:2011yk}; the blue shaded bands give the world average constraints on $a_{\rm sl}^d$ and $a_{\rm sl}^s$ individually; the red ellipse is the world average including all constraints~\cite{Amhis:2012bh}.
}
\label{fig:qovp}
\end{figure}

Nonetheless, it must be reiterated that there are several important observables that are not yet well measured, and that could rule out MFV.
For example, the bounds on NP scales obtained above (from Ref.~\cite{Isidori:2010kg}) do not include results on \CP violation in mixing in the \Bd and \Bs sectors.
In fact, the D0 collaboration has reported a measurement of an anomalous effect ~\cite{Abazov:2011yk} of the inclusive same-sign dimuon asymmetry, which is $3.9\sigma$ away from the SM prediction (of very close to zero~\cite{Lenz:2011ti}). 
This measurement is sensitive to an approximately equal combination of the parameters of \CP violation in \Bd and \Bs mixing, $a_{\rm sl}^d$ and $a_{\rm sl}^s$,\footnote{
  \label{foot-asl}
  The $a_{\rm sl}$ parameters, so named because the asymmetries are measured using semileptonic decays, are related to the $p$ and $q$ parameters by $a_{\rm sl} = (1 - \left|q/p\right|^4)/(1+\left|q/p\right|^4)$. 
} however some sensitivity to the source of the asymmetry can be obtained by applying additional constraints on the impact parameter to obtain a sample enriched in either oscillated $\Bd$ or $\Bs$ candidates. 
In addition, $a_{\rm sl}^d$ and $a_{\rm sl}^s$ can be measured individually.
The latest world average, shown in Fig.~\ref{fig:qovp}, gives
$a_{\rm sl}^d = -0.0003 \pm 0.0021$,
$a_{\rm sl}^s = -0.0109 \pm 0.0040$~\cite{Amhis:2012bh}.
Improved measurements are needed to resolve the situation.

\section{Current experimental status of heavy quark flavour physics}

\subsection{The CKM matrix and the Unitarity Triangle}

Much of the experimental programme in heavy quark flavour physics is devoted to measurements of the parameters of the CKM matrix.
As discussed above, the CKM matrix can be written in terms of the Wolfenstein parameters, which exploit the observed hierarchy in the mixing angles:
\begin{eqnarray}
  \label{eq:ckm}
  V_{\rm CKM} & = & 
  \left(
  \begin{array}{ccc}
    V_{ud} & V_{us} & V_{ub} \\
    V_{cd} & V_{cs} & V_{cb} \\
    V_{td} & V_{ts} & V_{tb} \\
  \end{array}
  \right) \nonumber \\
  & = &
  \left(
  \begin{array}{ccc}
    1 - \lambda^2/2 & \lambda & A \lambda^3 ( \rho - i \eta ) \\
    - \lambda & 1 - \lambda^2/2 & A \lambda^2 \\
    A \lambda^3 ( 1 - \rho - i \eta ) & - A \lambda^2 & 1 \\
  \end{array}
  \right) + {\cal O}\left( \lambda^4 \right) \, ,
\end{eqnarray}
where the expansion parameter $\lambda$ is the sine of the Cabibbo angle ($\lambda = \sin \theta_{\rm C} \approx V_{us}$).
It should be noted that although the hierarchy is highly suggestive, there is no underlying reason known for this pattern; moreover, the pattern in the lepton sector is completely different.
Note also that at ${\cal O}(\lambda^3)$ in the Wolfenstein parametrisation, the complex phase in the CKM matrix enters only in the $V_{ub}$ and $V_{td}$ (top right and bottom left) elements, but this is purely a matter of convention -- only relative phases are observable.  

The unitarity of the CKM matrix, $V^\dagger_{\rm CKM}V_{\rm CKM} = V_{\rm CKM}V^\dagger_{\rm CKM} = 1$, puts a number of constraints on the magnitudes and relative phases of the elements.
Among these relations, one which has been precisely tested is
\begin{equation}
  \left| V_{ud} \right|^2 + \left| V_{us} \right|^2 + \left| V_{ub} \right|^2 = 1 \, ,
\end{equation}
where the measurements of $\left| V_{ud} \right|^2$ from, \eg, super-allowed $\beta$ decays and $\left| V_{us} \right|^2$ from leptonic and semileptonic kaon decays are indeed consistent with the prediction to within one part in $10^3$~\cite{thetaCreview}.\footnote{
  The contribution from $\left| V_{ub} \right|^2$ is at the level of $10^{-5}$ and therefore negligible for this test at current precision.
}

\begin{figure}[!htb]
\centering
\includegraphics[width=0.40\textwidth]{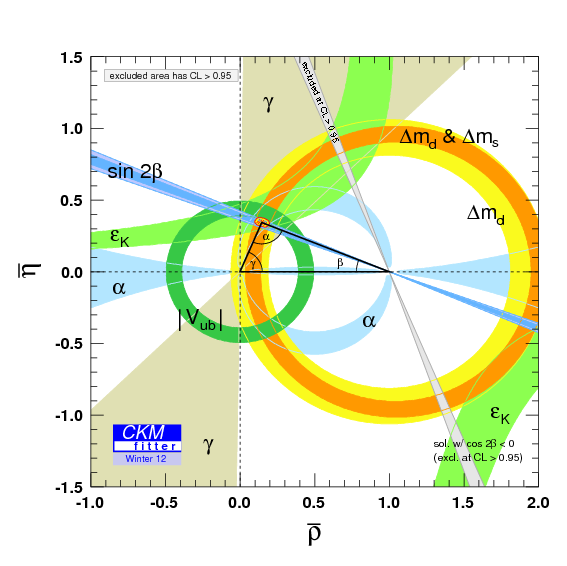}
\hspace{3mm}
\includegraphics[width=0.41\textwidth]{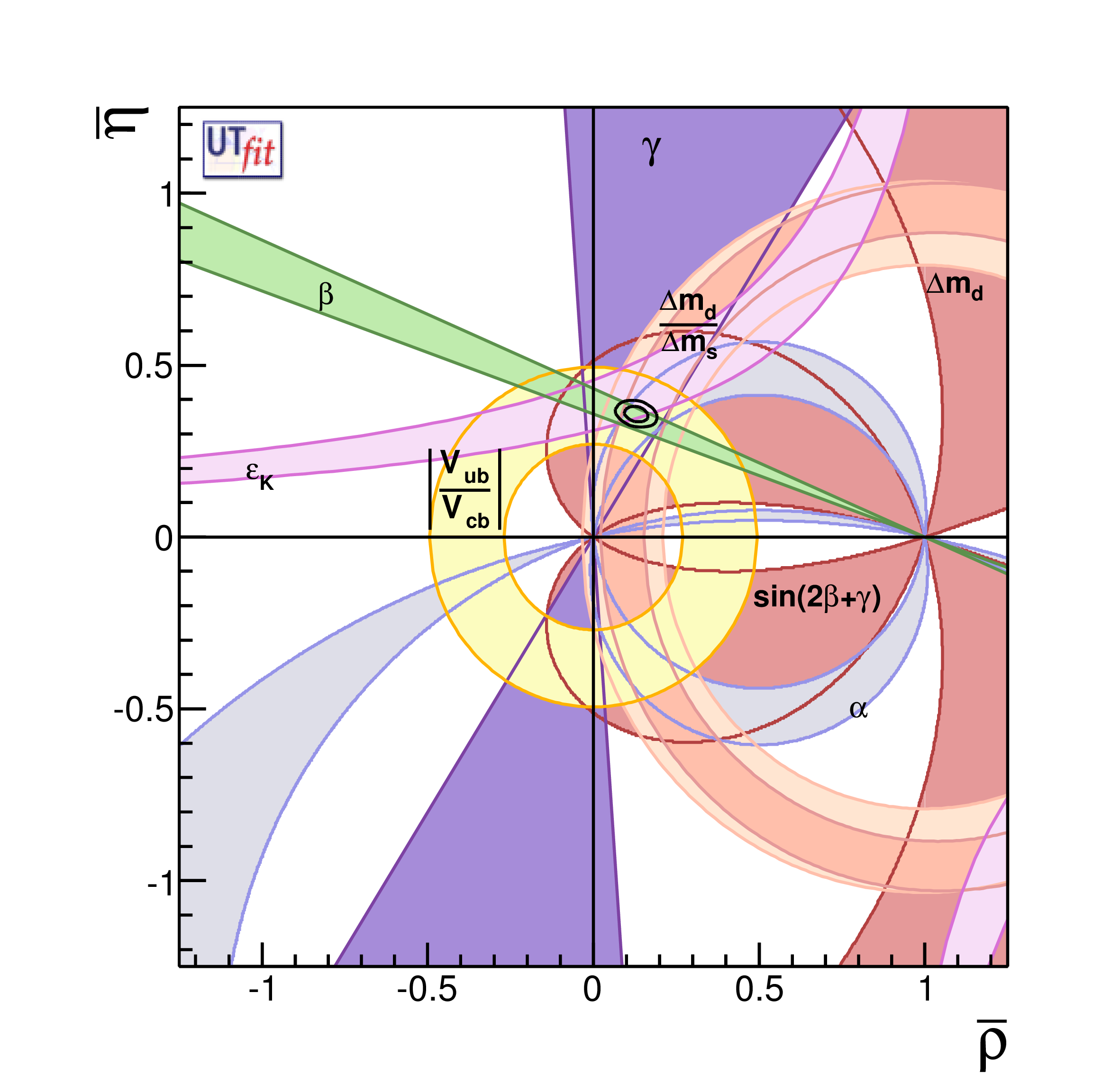}
\caption{\small
  Constraints on the Unitarity Triangle as compiled by (left) CKMfitter~\cite{Charles:2004jd}, (right) UTfit~\cite{Bona:2005vz}.
}
\label{fig:ckmfit}
\end{figure}

The unitarity condition also results in six constraints, $\Sigma_i V_{u_id_j}V_{u_id_k}^* = \Sigma_i V_{u_jd_i}V_{u_kd_i}^* = 0$ ($u_{i,j,k} \in (u,c,t),\, d_{i,j,k} \in (d,s,b),\, j\neq k$), for example
\begin{equation}
  \label{eq:ut}
  V_{ud} V_{ub}^* + V_{cd}V_{cb}^* + V_{td}V_{tb}^* = 0 \, ,
\end{equation}
which correspond to three complex numbers summing to zero, and hence can be represented as triangles in the complex plane.
The triangles have very different shapes, but all of them have the same area, which is given by half of the Jarlskog parameter~\cite{Jarlskog:1985ht}.
The specific triangle relation given in Eq.~(\ref{eq:ut}) is particularly useful to visualise the constraints from various different measurements, as shown in the iconic images from the CKMfitter~\cite{Charles:2004jd} and UTfit~\cite{Bona:2005vz} collaborations, reproduced in Fig.~\ref{fig:ckmfit}.
Conventionally, this ``Unitary Triangle'' (UT) is rescaled by $V_{cd}V_{cb}^*$ so that by definition the position of the apex is
\begin{equation}
  \bar{\rho} + i \bar{\eta} \equiv - \frac{V_{ud} V_{ub}^*}{V_{cd}V_{cb}^*} \, ,
\end{equation}
where $\left( \rhobar, \etabar \right)$~\cite{Buras:1994ec} are related to the Wolfenstein parameters by
\begin{equation}
  \label{eq:rhoetabarinv}
  \rho + i\eta \;=\; 
  \frac{ 
    \sqrt{ 1-A^2\lambda^4 }(\bar{\rho}+i\bar{\eta}) 
  }{
    \sqrt{ 1-\lambda^2 } \left[ 1-A^2\lambda^4(\bar{\rho}+i\bar{\eta}) \right]
  } \, .
\end{equation}

Two popular naming conventions for the UT angles exist in the literature:
\begin{equation}
  \label{eq:ut-angles}
  \alpha  \equiv  \phi_2  = 
  \arg\left[ - \frac{V_{td}V_{tb}^*}{V_{ud}V_{ub}^*} \right],
  \hspace{0.2cm}
  \beta   \equiv   \phi_1 =  
  \arg\left[ - \frac{V_{cd}V_{cb}^*}{V_{td}V_{tb}^*} \right],
  \hspace{0.2cm}
  \gamma  \equiv   \phi_3  =  
  \arg\left[ - \frac{V_{ud}V_{ub}^*}{V_{cd}V_{cb}^*} \right].
\end{equation}
The $\left( \alpha, \beta, \gamma \right)$ set is used in these lectures.
The lengths of the sides $R_u$ and $R_t$ of the UT are given by
\begin{equation}
  \label{eq:ru_rt}
  R_u =
  \left|\frac{V_{ud}V_{ub}^*}{V_{cd}V_{cb}^*} \right|
  = \sqrt{\rhobar^2+\etabar^2}\,,
  \hspace{0.5cm}
  R_t = 
  \left|\frac{V_{td}V_{tb}^*}{V_{cd}V_{cb}^*}\right| 
  = \sqrt{(1-\rhobar)^2+\etabar^2}\,.
\end{equation} 

A major achievement of the past decade or so has been to significantly improve the precision of the parameters of the UT.
In particular, the primary purpose of the so-called ``$B$ factory'' experiments, BaBar and Belle, was the determination of $\sin 2\beta$ using $B^0 \to \jpsi \KS$ (and related modes).
This was carried out using completely new experimental techniques to probe \CP violation in a very different way to previous experiments in the kaon system.
In particular, if we denote the amplitude for a $\Bz$ meson to decay to a particular final state $f$ as $A_f$, and that for the charge conjugate process as $\bar{A}_{\bar{f}}$, then using the parameters $p$ and $q$ from Eq.~(\ref{eq:pqdef}), we define the parameter $\lambda_f = \frac{q}{p}\frac{\bar{A}_{\bar{f}}}{A_f}$ and the following categories of \CP violation in hadronic systems:\footnote{
  Considering the possibility that \CP violation may be observed in the lepton sector as differences of oscillation parameters between neutrinos and antineutrinos (in appearance experiments), it is worth noting that this would be another different category.
}
\begin{enumerate}
\item \CP violation in mixing ($\left| q/p \right| \neq 1$),
\item \CP violation in decay ($\left| \bar{A}_{\bar{f}}/A_f \right| \neq 1$),
\item \CP violation in interference between mixing and decay (${\rm Im}\left( \lambda_f \right) \neq 0$).
\end{enumerate}
Additionally, in the literature the concepts of {\it indirect} and {\it direct} \CP violation are often encountered: the former is where the effect is consistent with originating from a single phase in the mixing amplitude, while the latter cannot be accounted for in such a way.
Following this categorisation, \CP violation in decay (the only category available to baryons or charged mesons) is direct, while \CP violation in mixing and interference can be indirect so long as only one measurement is considered -- but if two such measurements give different values, this also establishes direct \CP violation.

\subsection{Determination of $\sin(2\beta)$}

The determination of $\sin(2\beta)$ from $\Bd \to \jpsi \KS$~\cite{Carter:1980tk,Bigi:1981qs}, exploits the fact that some measurements of \CP violation in interference between mixing and decay can be cleanly interpreted theoretically, since hadronic factors do not contribute. 
The full derivation of the decay-time-dependent decay rate of $\Bz$ mesons to a \CP eigenstate $f$ is a worthwhile exercise for the reader, and can be found in, \eg, Refs.~\cite{Branco:1999fs,Bigi:2000yz}.
The result, for mesons that are known to be either $\Bzb$ or $\Bz$ at time $t=0$, is
\begin{eqnarray}
  \Gamma(\Bzb_{\rm phys} \to f(t)) & = &
  \frac{e^{-t/\tau(\Bz)}}{2\tau(\Bz)}
  \left[ 1 + S_f \sin(\Delta m t) - C_f \cos(\Delta m t) \right], \nonumber \\
  \Gamma(\Bz_{\rm phys} \to f(t)) & = &
  \frac{e^{-t/\tau(\Bz)}}{2\tau(\Bz)}
  \left[ 1 - S_f \sin(\Delta m t) + C_f \cos(\Delta m t) \right],
  \label{eq:SandC}
\end{eqnarray}
where 
\begin{equation}
  S_f = \frac{2\, {\rm Im}(\lambda_f)}{1 + |\lambda_f|^2}\,,
  \ {\rm and} \
  C_f = \frac{1 - |\lambda_f|^2}{1 + |\lambda_f|^2}\, .
\end{equation}
In these expressions $\Delta \Gamma$ has been assumed to be negligible, as appropriate for the $\Bd$ system.
Assuming further $\left| q/p \right| = 1$, then for decays dominated by a single amplitude, $C_f = 0$ and $S_f = \sin ({\rm arg}(\lambda_f) )$, and so for $\Bd \to \jpsi \KS$, $S = \sin(2\beta)$, to a very good approximation.

The experimental challenge for the measurement of $\sin(2\beta)$ then lies in the ability to measure the coefficient of the sinusoidal oscillation of the decay-time-asymmetry.
Until recently, the most copious sources of cleanly reconstructed $B$ mesons came from accelerators colliding electrons with positrons at the $\Upsilon(4S)$ resonance (a $b\bar{b}$ bound state just above the threshold for decay into pairs of $B$ mesons).
For symmetric colliders, the $B$ mesons are produced at rest, and therefore lifetime measurements are not possible. 
A boost is necessary, which can be advantageously achieved by making the $e^+e^-$ collisions asymmetric.\footnote{
  Boosted $b$ hadrons can also be obtained in hadron colliders, as will be discussed below.
}
One strong feature of this approach is that the quantum correlations of the $B$ mesons produced in $\Upsilon(4S)$ decay are retained, so that the decay of one into a final state that tags its flavour ($\Bz$ or $\Bzb$) can be used to set the clock to $t=0$ and specify the flavour of the other at that time.

The concept of the asymmetric $B$ factory was such a good one that two were built: PEP-II at SLAC, colliding $9.0 \gev$ $e^-$ with $3.1 \gev$ $e^+$, and KEKB at KEK ($8.0 \gev$ $e^-$ on $3.5 \gev$ $e^+$).
These have achieved world record luminosities, with peak instantaneous luminosities above $10^{34} \cm^{-2} \sec^{-1}$, and a combined data sample of over $1 \invab$, corresponding to over $10^9$ $B\bar{B}$ pairs.
The detectors (BaBar~\cite{Aubert:2001tu,BABAR:2013jta} and Belle~\cite{Abashian:2000cg} respectively) share many common features, such as silicon vertex detectors, gas based drift chambers, electromagnetic calorimeters based on Tl-doped CsI crystals, and $1.5 \, {\rm T}$ solenoidal magnetic fields.
The main difference is in the technology used to separate kaons from pions: a system based on the detection of internally reflected Cherenkov light for BaBar, and a combination of aerogel Cherenkov counters and a time-of-flight system for Belle.

\begin{figure}[!htb]
\centering
\includegraphics[width=0.40\textwidth]{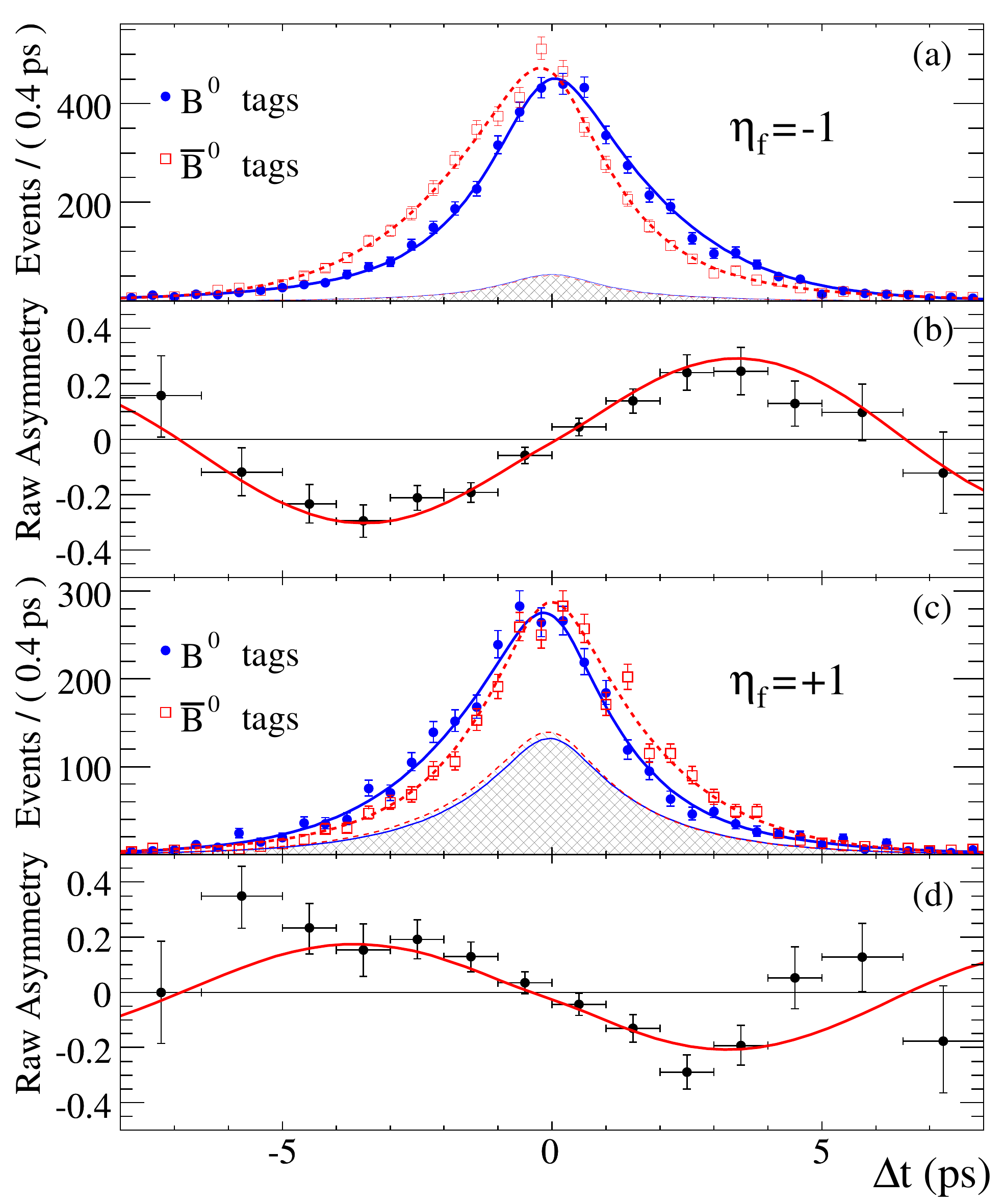}
\hspace{3mm}
\includegraphics[width=0.265\textwidth]{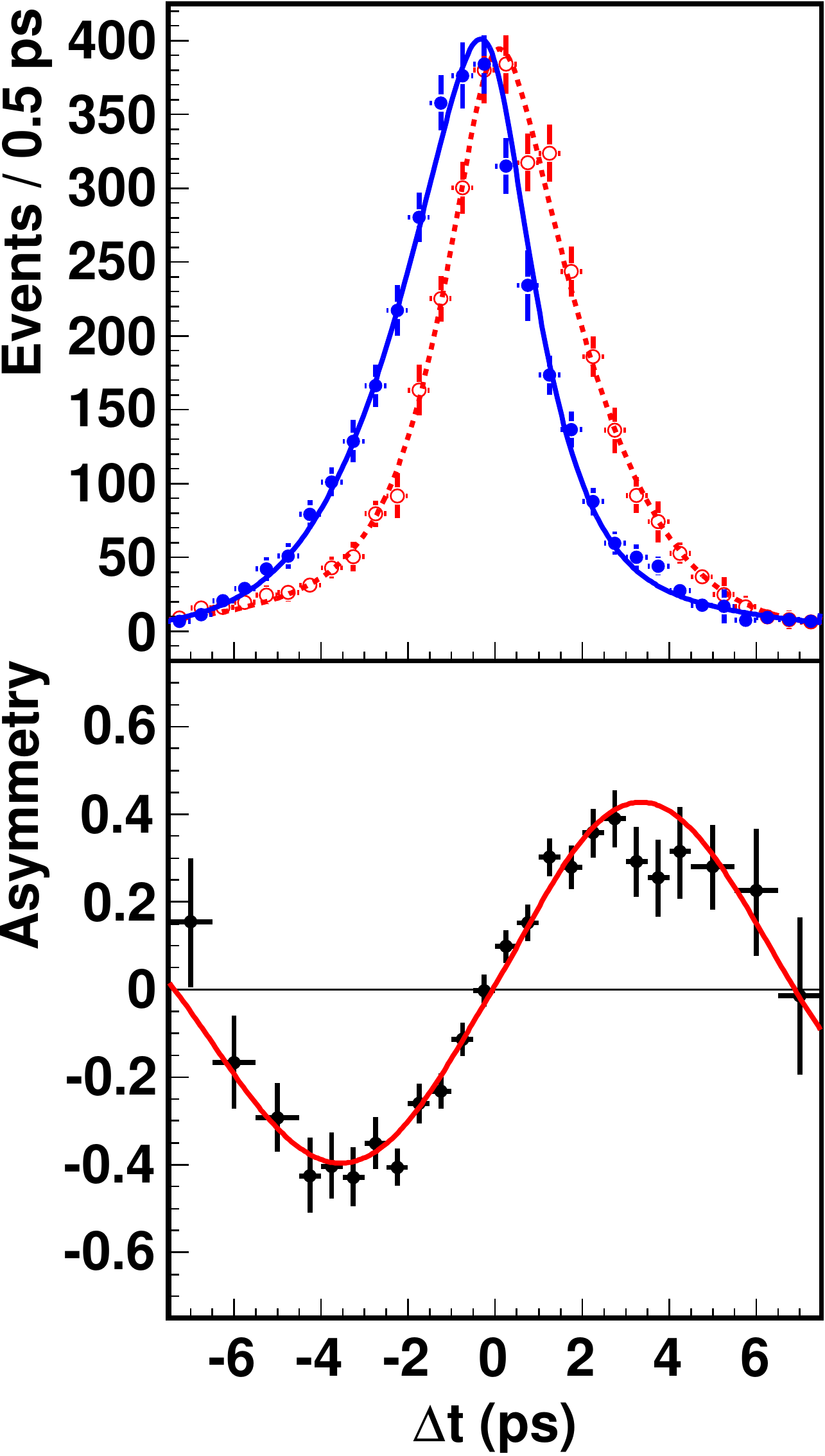}
\caption{\small
  Results from (left) BaBar~\cite{Aubert:2009aw} and (right) Belle~\cite{Adachi:2012et} on the determination of $\sin2\beta$.  
}
\label{fig:sin2beta}
\end{figure}

Through the measurement of $\sin 2\beta$, BaBar~\cite{Aubert:2001nu} and Belle~\cite{Abe:2001xe} were able to make the first observations of \CP violation outside the kaon sector, thus validating the Kobayashi-Maskawa mechanism.
The latest (and, excluding upgrades, most likely final) results from BaBar~\cite{Aubert:2009aw} and Belle~\cite{Adachi:2012et} shown in Fig.~\ref{fig:sin2beta} give a clear visual confirmation of the large \CP violation effect.
The world average value, using determinations based on $b \to c\bar{c}s$ transitions, is~\cite{Amhis:2012bh}
\begin{equation}
  \sin 2 \beta = 0.682 \pm 0.019 \ {\rm which \ gives} \ \beta = (21.5 \,^{+0.8}_{-0.7})^\circ \, .
\end{equation}

\subsection{Determination of $\alpha$}

Additional measurements are needed to over-constrain the UT and thereby test the Standard Model.
The angle $\alpha$ can, in principle, be determined in a similar way as $\beta$, but using a decay mediated by the $b \to u\bar{u}d$ tree-diagram which carries the relative weak phase $\gamma$ (since $\pi - \left( \beta + \gamma \right) = \alpha$ by definition).
However, in any such decay a contribution from the $b \to d$ loop (``penguin'') amplitude, which carries a different weak phase, is also possible. 
This complicates the interpretation of the observables, since $S \neq \sin(2\alpha)$; on the other hand direct \CP violation becomes observable, if the relative strong phase is non-zero.
Constraints on $\alpha$ can still be obtained using a channel in which the penguin contribution either can be shown to be small, or can be corrected for using an isospin analysis~\cite{Gronau:1990ka}.
The world average, $\alpha = \left( 89.0 \,^{+4.4}_{-4.2} \right)^\circ$, is dominated by constraints from the $\Bz \to \rho^+\rho^-$ decay~\cite{Aubert:2007nua,Abe:2007ez}, which is consistent with having negligible penguin contribution.

\subsection{The sides of the Unitarity Triangle}

The lengths of the sides of the UT have also been constrained by various observables.
The value of $R_t$ depends on $\left| V_{td}  \right|$, and can be determined from $b \to d$ transitions such as the rate of $\Bd$ oscillations, \ie\ $\Delta m_d$, or the branching fraction $B \to \rho \gamma$.  
In both cases, theoretical uncertainties are reduced if the measurement is performed relative to that for the corresponding $b \to s$ transition.
The most precise constraint to date comes from the ratio of $\Delta m_d$~\cite{Amhis:2012bh,Abe:2004mz,LHCb-PAPER-2012-032} and $\Delta m_s$~\cite{Amhis:2012bh,Abulencia:2006ze,LHCb-PAPER-2013-006} and gives $\left| \frac{V_{td}}{V_{ts}} \right| = 0.211 \pm 0.001 \pm 0.005$, where the first uncertainty is experimental and the second theoretical (originating from lattice QCD calculations).

The value of $R_u$ depends on $\left| V_{ub} \right|$ and can be determined from $b \to u$ tree-level transitions. 
Semileptonic decays allow relatively clean theoretical interpretation,\footnote{
  Fully leptonic decays are even cleaner theoretically, but are experimentally scarce.  Such modes will be discussed below.
}
but still require QCD calculations to go from the parton level transition to the observed (semi-hadronic) final state (for a recent review, see Ref.~\cite{Luth:2011zz}).
Two approaches have been pursued: exclusive decays, such as $\Bz \to \pi^- e^+\nu$, and inclusive decays, $B \to X_u e^+\nu$.
The theory of inclusive decays is based on the operator product expansion (discussed in Sec.~\ref{sec:kstarmumu}) and would be extremely clean, were it not for the fact that experimentally cuts are needed to remove the more prevalent $b \to c$ transition.
Exclusive decays tend to have less background from $b \to c$ processes.
The differential branching fractions can be translated in constraints on $\left| V_{ub} \right|$ using knowledge of form-factors at the kinematic limit obtained from lattice QCD calculations, together with phenomenological models that extrapolate over the whole phase space.
The most precise results use $B \to \pi \ell^+\nu$ decays ($\ell = e,\mu$)~\cite{delAmoSanchez:2010zd,delAmoSanchez:2010af,Ha:2010rf}, and give an ``exclusive'' determination of $\left| V_{ub} \right|$ that is, however, in tension with the ``inclusive'' value~\cite{Amhis:2012bh}:
\begin{eqnarray*}
  \left| V_{ub} \right|_{\rm excl.} & = & \left[ 3.23 \times \left( 1.00 \pm 0.05 \pm 0.08 \right) \right] \times 10^{-3} \, ,\\
  \left| V_{ub} \right|_{\rm incl.} & = & \left[ 4.42 \times \left( 1.000 \pm 0.045 \pm 0.034 \right) \right] \times 10^{-3} \, .
\end{eqnarray*} 
where the first uncertainties are experimental and the second theoretical.
Since the origin of the discrepancy, which is also seen in determinations of $\left| V_{cb} \right|$ from $b \to c \ell\nu$ transitions, is not understood, the uncertainty is scaled to give 
\begin{eqnarray*}
  \left| V_{ub} \right|_{\rm avg.} & = & \left[ 3.95 \times \left( 1.000 \pm 0.096 \pm 0.099 \right) \right] \times 10^{-3} \, .
\end{eqnarray*}

The results on $\beta$, $\alpha$, $R_t$ and $R_u$ are the most constraining inputs to the CKM fits shown in Fig.~\ref{fig:ckmfit}~\cite{Charles:2004jd,Bona:2005vz}.
While the results are all consistent with the Standard Model prediction of a single source of \CP violation, there are some tensions that deserve further investigation.
Moreover, there are still certain important observables where large NP contributions are possible, as will be discussed in more detail below.

\section{Flavour physics at hadron colliders}

Results from the $B$ factory experiments provided enormous progress in the understanding of heavy flavour physics (only a very brief selection has been discussed above).
Nonetheless, many results remain statistically limited, and the $\Bs$ sector is relatively unexplored.
To progress further, it is necessary to have a copious source of production of all flavours of $b$ hadron. 
As shown in Table~\ref{tab:gibson}, high energy hadron colliders satisfy these criteria, but present significant experimental challenges: to be able to identify the decays of interest from the high multiplicity environment, and to reject the even more copious rate of minimum bias events.\footnote{
  Experiments at $e^+e^-$ machines also have to reject effectively backgrounds from QED processes, but this can be done at trigger level with simple requirements.
} 

\begin{table}[!htb]
  \caption{Summary of some relevant properties for $b$ physics in different experimental environments.  Adapted from Ref.~\cite{GibsonHCPSS}.
  }
  \label{tab:gibson}
  \centering
\begin{tabular}{lccc}
  \hline\noalign{\smallskip}
  & $e^+e^- \to \Upsilon(4S) \to B\bar{B}$ & $p\bar{p} \to  b\bar{b}X$ & $pp \to  b\bar{b}X$ \\
  & & ($\sqrt{s} = 2 \tev$) & ($\sqrt{s} = 14 \tev$) \\
  & PEP-II, KEKB & Tevatron & LHC \\
\noalign{\smallskip}\svhline\noalign{\smallskip}
  Production cross-section & $1 \nb$ & $\sim 100 \mub$ & $\sim 500 \mub$ \\
  Typical $b\bar{b}$ rate & $10 \hz$ & $\sim 100 \khz$ & $\sim 500 \khz$ \\
  Pile-up & 0 & 1.7 & 0.5--20 \\
  $b$ hadron mixture & $\Bp\Bm$ (50\%), $\Bz\Bzb$ (50\%) & \multicolumn{2}{c}{\Bp (40\%), \Bz (40\%), \Bs (10\%),} \\
  & & \multicolumn{2}{c}{$\Lb$ (10\%), others ($<1\%$)} \\
  $b$ hadron boost & small ($\beta\gamma \sim 0.5$) & \multicolumn{2}{c}{large ($\beta\gamma \sim 100$)} \\ 
  Underlying event & $B\bar{B}$ pair alone & \multicolumn{2}{c}{Many additional particles} \\
  Production vertex & Not reconstructed & \multicolumn{2}{c}{Reconstructed from many tracks} \\
  $\Bz$--$\Bzb$ pair production & Coherent (from $\Upsilon(4S)$ decay) & \multicolumn{2}{c}{Incoherent} \\
  Flavour tagging power & $\epsilon D^2 ~ \sim 30\%$ & \multicolumn{2}{c}{$\epsilon D^2 ~ \sim 5\%$} \\
\noalign{\smallskip}\hline\noalign{\smallskip}
\end{tabular}
\end{table}

\begin{figure}[!htb]
\centering
\includegraphics[width=0.8\textwidth]{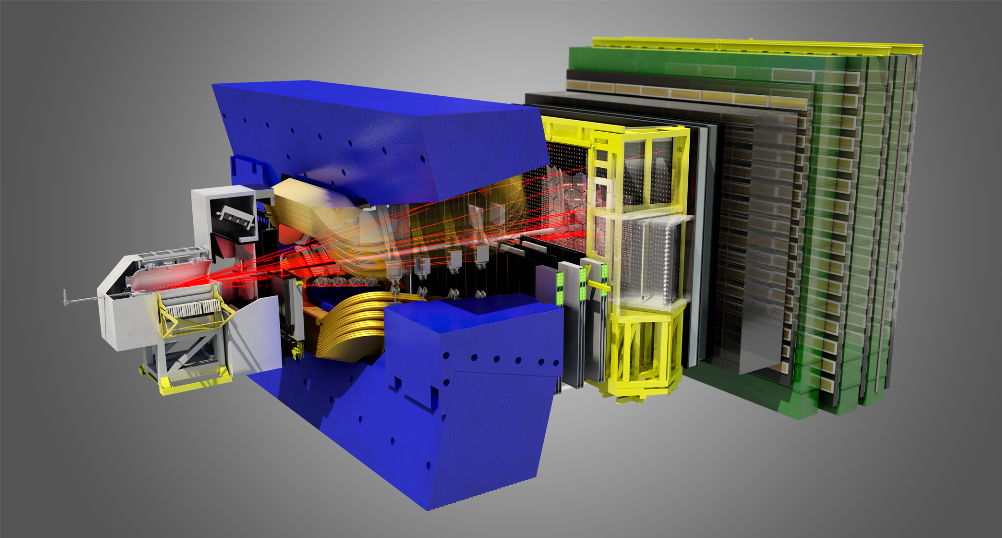}
\caption{
  The LHCb detector~\cite{Alves:2008zz}.
}
\label{fig:lhcb}
\end{figure}

The LHCb detector~\cite{Alves:2008zz}, shown in Fig.~\ref{fig:lhcb}, has been designed to meet these challenges.
It is in essence a forward spectrometer (covering the acceptance region that optimises its flavour physics capability), with a dipole magnet, a precision silicon vertex detector and strong particle identification capability.
Tracks can be identified as different hadron species using information from ring-imaging Cherenkov detectors, while calorimeters and muon detectors enable charged leptons to be distinguished and also provide trigger signals.
The trigger system~\cite{LHCb-DP-2012-004} uses these hardware level signals to reduce the rate from the maximum LHC bunch-crossing rate of $40 \mhz$ to the $1 \mhz$ rate at which the detector can be read out.
A software trigger then searches for inclusive signatures of $b$-hadron decays such as high-\pt signals and displaced vertices, and also performs reconstruction of several exclusive $b$ and $c$ decay channels, in order to further reduce the rate to a level that can be written to offline data storage ($3\khz$ in 2011, $5 \khz$ in 2012).

During the LHC run, the detector operated with data taking efficiency above $90\%$, with instantaneous luminosity around $3\,(4) \times 10^{32} \cm^{-2} \sec^{-1}$ recording data samples of $1 \, (2) \invfb$ at $\sqrt{s} = 7 \,(8) \tev$ in 2011 (2012).\footnote{
  Note that these values already exceed the LHCb design luminosity of $2 \times 10^{32} \cm^{-2} \sec^{-1}$.
}
The luminosity is less than that delivered to ATLAS and CMS, since the experimental design requires low pile-up, \ie\ a low number of $pp$ collisions per bunch-crossing.
However, this allows the luminosity to be ``levelled'' and remain at a constant value throughout the LHC fill, providing very stable data taking-conditions.\footnote{
  Similar stability was achieved at $e^+e^-$ colliders by a completely different method referred to as trickle (or continuous) injection.
}
In addition, the polarity of the dipole magnet is reversed regularly, which enables cancellation of detector asymmetries in various measurements.

In addition to LHCb, it must be noted that the ``general purpose detectors'' ATLAS and CMS at the LHC, and CDF and D0 at the Tevatron, have capability to study flavour physics.
For most of these experiments, their programme is, however, restricted to decay modes triggered by high \pt muons, but CDF benefited from a two-track trigger~\cite{Ristori:2010zz} that enabled a broader range of measurements to be performed.

\subsection{Heavy flavour production and spectroscopy}

The capabilities of the different experiments can be demonstrated from the measurements of production cross-sections that have been performed by each.
Most have studied \jpsi production (\eg\ Refs.~\cite{Acosta:2004yw,Aad:2011sp,Chatrchyan:2011kc,LHCb-PAPER-2011-003,LHCb-PAPER-2012-039,LHCb-PAPER-2013-016}) as well as $b$ hadron production using decay modes involving muons or \jpsi mesons~\cite{Aaltonen:2007zza,Aaltonen:2009xn,Abachi:1994kj,Aad:2012jga,Chatrchyan:2012hw,LHCb-PAPER-2010-002}.
However, only CDF and LHCb have been able to study prompt charm production~\cite{Acosta:2003ax,LHCb-PAPER-2012-041}.\footnote{
  Measurements of charm production and other processes by ALICE are not included in this discussion.  Although ALICE can study production at low luminosity, it cannot perform competitive studies of flavour changing processes.
}
The cross-sections measured confirm the theoretical predictions, and enable the values for integrated luminosity to be translated into more easily comprehensible terms.
For example, with $1 \invfb$ recorded at $\sqrt{s} = 7 \tev$, and the measured $b\bar{b}$ production cross-section~\cite{LHCb-PAPER-2010-002,LHCb-PAPER-2011-018}, it is easily shown that over $10^{11}$ $b\bar{b}$ quark pairs have been produced in the LHCb acceptance.  
This compares to the combined BaBar and Belle data sample of $\sim 10^{9}$ $B\bar{B}$ meson pairs. 
Consequently, for any channel where the efficiency, including effects from reconstruction, trigger and offline selection, is not too small, LHCb has the world's largest data sample. 
This further emphasises the need for an excellent trigger in order to perform flavour physics at hadron colliders.

Production measurements such as those mentioned above test QCD models, and are important (and highly-cited) results.
However, since they are not within the remit of flavour-changing interactions of the charm and beauty quarks, they will not be discussed further here.
Nonetheless, a brief digression into studies of another aspect of QCD, that of spectroscopy, will be worthwhile.
This covers the study of properties of hadronic states such as lifetimes, masses, decay channels and quantum numbers, and also the discoveries of new states.
Indeed, some of the most highly-cited papers from recent flavour physics experiments relate to such topics, including the discovery of the $X(3872)$ particle by Belle~\cite{Choi:2003ue} and of the $D_{sJ}$ states by BaBar~\cite{Aubert:2003fg} and CLEO~\cite{Besson:2003cp}.
The first new particles discovered at the LHC, prior to the Higgs boson, were hadronic states~\cite{Aad:2011ih,Chatrchyan:2012ni,LHCB-PAPER-2012-012}.
More recently, significant progress has been made in understanding the nature of the $X(3872)$~\cite{LHCb-PAPER-2013-001}.
New results are eagerly anticipated in several related areas, for example to clarify the situation regarding the existence of charged charmonium-like states, claimed by Belle~\cite{Choi:2007wga,Mizuk:2009da,Mizuk:2008me} but not confirmed by BaBar~\cite{Aubert:2008aa,Lees:2011ik}, which would be smoking gun signatures for an exotic hadronic nature.\footnote{
  New claims of charged charmonium-like states have recently been made~\cite{Ablikim:2013mio,Liu:2013dau}.
}
Recent claims of charged bottomonium-like states by Belle~\cite{Belle:2011aa,Adachi:2012cx} seem to strengthen the case that such exotics can exist in nature, but one should note that history teaches us that not all claimed states turn out to be real~\cite{Hom:1976cv}.

The topic of spectroscopy also provides a useful illustration of the importance of triggering for flavour physics experiments at hadron colliders.
In 2008, the BaBar experiment discovered the $\eta_b$ meson using the process $e^+e^- \to \Upsilon(3S) \to \eta_b \gamma$, where only the photon is reconstructed and the signal is inferred from a peak in the photon energy spectrum~\cite{Aubert:2008ba}.
The $\eta_b$ meson is the pseudoscalar $b\bar{b}$ ground state.
It is the lightest bottomonium state, so why did it take more than 30 years after the discovery of the $\Upsilon(1S)$ meson~\cite{Herb:1977ek} (the lightest vector $b\bar{b}$ state) to see it in experiments? 
In particular, since hadron collisions produce particles with all possible quantum numbers, why was it not observed at, \eg, the Tevatron?
The answer lies in the fact that the vector state decays to dimuons, which have a distinctive trigger signature.  
The dominant decay channels of the $\eta_b$ are expected to be multibody hadronic final states, which make its observation in a hadronic environment extremely challenging.

\section{Key observables in the LHC era}

\subsection{Direct \CP violation}
\label{subsec:dcpv}

\begin{figure}[!htb]
\centering
\includegraphics[width=0.45\textwidth]{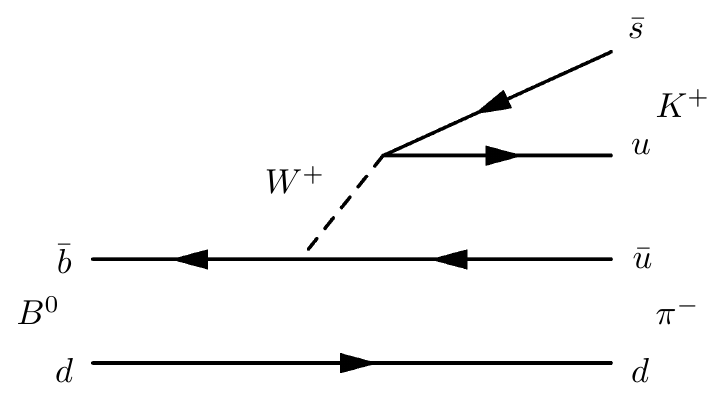}
\includegraphics[width=0.45\textwidth]{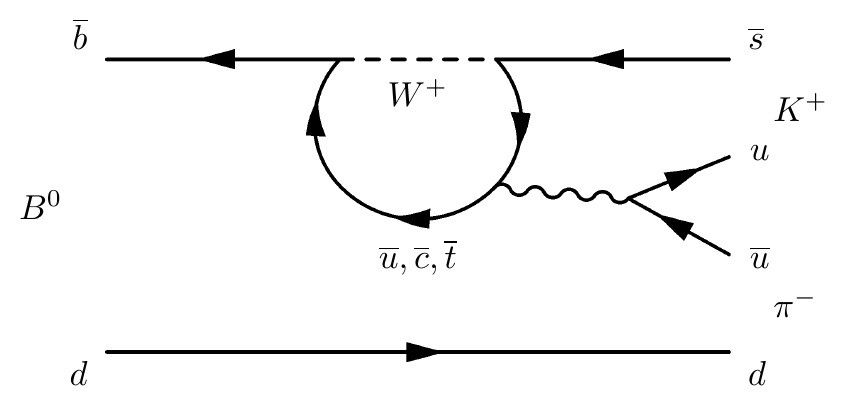}
\caption{
  SM (left) tree and (right) penguin diagrams for the decays $\Bz\to\Kp\pim$.
}
\label{fig:TandP}
\end{figure}

As mentioned above, a condition for direct \CP violation is $\left|\bar{A}_{\bar{f}}/A_f \right| \neq 1$.
In order for this to be realised we need the amplitude to be composed of at least two parts with different weak and strong phases.
This is often realised by tree ($T$) and penguin ($P$) amplitudes (example diagrams are shown in Fig.~\ref{fig:TandP}), so that
\begin{equation}
  A_f = \left| T \right| e^{i(\delta_T - \phi_T)} + \left| P \right| e^{i(\delta_P -\phi_P)} \ \ {\rm and} \ \
  \bar{A}_{\bar{f}} = \left| T \right| e^{i(\delta_T + \phi_T)} + \left| P \right| e^{i(\delta_P + \phi_P)} \, ,
\end{equation}
where the strong (weak) phases $\delta_{T,P}$ ($\phi_{T,P}$) keep the same (change) sign under the \CP transformation by definition.
The \CP asymmetry is defined from the rate difference between the particle involving the quark ($D$ or $\bar{B}$) and that containing the antiquark ($\bar{D}$ or $B$).
Using the definition for $B$ decays, we trivially find
\begin{equation}
  A_{\CP} = \frac{\left| \bar{A}_{\bar{f}} \right|^2 - \left| A_{f} \right|^2}{\left| \bar{A}_{\bar{f}} \right|^2 + \left| A_{f} \right|^2} =
  \frac{2 \left| T \right| \left| P \right| \sin(\delta_T - \delta_P) \sin(\phi_T - \phi_P)} {\left| T \right|^2 + \left| P \right|^2 + 2 \left| T \right| \left| P \right| \cos(\delta_T - \delta_P) \cos(\phi_T - \phi_P)} \, .
\end{equation}
Therefore, for large direct \CP violation effects to occur, we need $\left| P/T \right|$, $\sin(\delta_T - \delta_P)$ and $\sin(\phi_T - \phi_P)$ to all be ${\cal O}(1)$.

Charmless $B$ decays, \ie\ decays of $B$ mesons to final states that do not contain charm quarks, provide good possibilities for the observation of direct \CP violation, since many decays have both tree and penguin contributions with similar levels of CKM suppression.
These are of interest to search for NP, since the penguin loop diagrams are sensitive to potential contributions from new particles.
An excellent example is $\Bd \to \Kp\pim$, which provided the first observation of direct \CP violation outside the kaon sector, and has a world average value of $A_{\CP}(\Bd \to \Kp\pim) = -0.086 \pm 0.007$~\cite{Lees:2012kx,Duh:2012ie,LHCb-PAPER-2012-029,CDFnote10726,Amhis:2012bh}.
Curiously, the \CP violation effect observed in $\Bp \to \Kp\piz$ decays is rather different: $A_{\CP}(\Bp \to \Kp\piz) = 0.040 \pm 0.021$~\cite{Lees:2012kx,Duh:2012ie,Amhis:2012bh}, although na\"ively changing the spectator quark in Fig.~\ref{fig:TandP} suggests that similar values should be expected.
This is referred to as the ``$K\pi$ puzzle'', and could in principle be a hint for NP, though the more mundane explanation of larger than expected QCD corrections cannot be ruled out at present.
Several methods are available to test the QCD explanations, which motivate improved measurements of other $K\pi$ modes (in particular, of $A_{\CP}(\Bd\to\KS\piz)$), of similar decay modes with three-body final states ($K\rho,\Kstar\pi$), and of charmless two-body \Bs decays.
On this last topic, following pioneering work by CDF~\cite{Aaltonen:2011jv,CDFnote10726}, LHCb has recently reported both the first decay time-dependent analysis of $\Bs \to \Kp\Km$~\cite{LHCb-CONF-2012-007} and the first observation of \CP violation in $\Bs \to \Km\pip$ decays~\cite{LHCb-PAPER-2013-018}, which demonstrate good prospects for progress in the coming years.

With regard to three-body decays, it is worth noting that despite hundreds of measurents by the $B$ factories, the significance of the world average in any other charmless $\Bp$ or $\Bz$ decay mode does not exceed $5\,\sigma$, though channels such as $\Bp \to \eta \Kp$ and $\Bp \to \rho^0 \Kp$ approach this level.
However, very recently, LHCb has demonstrated that large \CP violation effects occur in specific regions of the phase space of three-body charmless decays such as $\Bp \to \Kp\pip\pim$~\cite{LHCb-CONF-2012-018,LHCb-CONF-2012-028,LHCb-PAPER-2013-027}.
Further study is necessary to quantify the effect and identify its origin.

\subsection{The UT angle $\gamma$ from $B \to DK$ decays}

The angle $\gamma$ of the CKM Unitarity Triangle is unique in that it is the only \CP-violating parameter that can be measured using only tree-level decays.
This makes it a benchmark Standard Model reference point.
Improving the precision with which $\gamma$ is known is one of the primary goals of contemporary flavour physics, and this will only become more important after NP is discovered, since it will be essential to disentangle SM and NP contributions to \CP-violating observables.

The phase $\gamma$ can be determined exploiting the fact that in decays of the type $B \to DK$, the $b \to c \bar{u}s$ and $b \to u \bar{c} s$ amplitudes can interfere if the neutral charmed meson is reconstructed in a final state that is accessible to both \Dz and \Dzb decays.
There are many possible such final states, with various experimental advantages and disadvantages.
These include \CP eigenstates, doubly- or singly-Cabibbo-suppressed decays and multibody final states.
Moreover, decays of different $b$ hadrons can all be used to provide constraints on $\gamma$.
Two particularly interesting approaches are to study decay time-dependent asymmetries of $\Bs\to\Dsmp\Kpm$ decays~\cite{Aleksan:1991nh} and to study the Dalitz plot (\ie\ phase-space) dependent asymmetries in $\Bd\to D\Kp\pim$ decays~\cite{Gershon:2008pe,Gershon:2009qc}.
First results from LHCb show promising potential for these decays~\cite{LHCb-CONF-2012-029,LHCb-PAPER-2013-022}.
All such measurements will help to improve the overall precision in a combined fit.

\begin{figure}[!htb]
\centering
\includegraphics[width=0.75\textwidth]{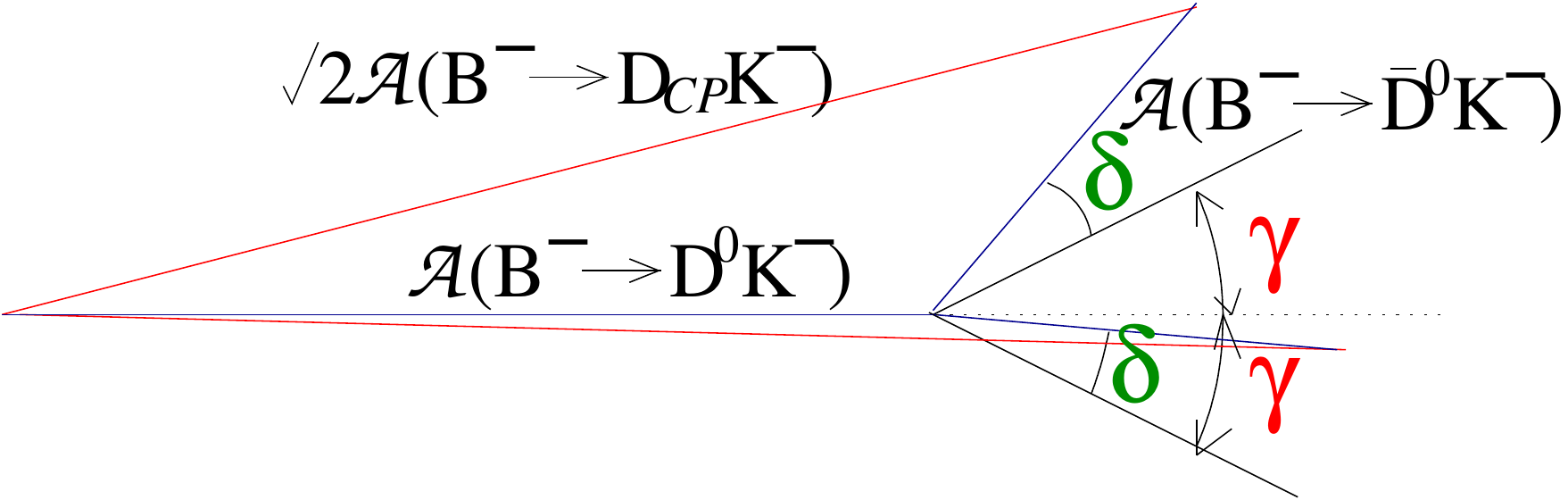}
\caption{
  Illustration of the concept behind the determination of $\gamma$ using $\Bpm \to D\Kpm$ decays.
  For \Bm decays the amplitudes add with relative phase $\delta - \gamma$, while for \Bp the relative phase is $\delta + \gamma$.
  Here the simplest case with \D decays to \CP eigenstates (such as $\Kp\Km$) is shown, but the method can be extended to any final state accessible to both \Dz and \Dzb decays.
}
\label{fig:gamma-method}
\end{figure}

The basic concept behind the method is illustrated in Fig.~\ref{fig:gamma-method} for $B^- \to D_{\CP} K^-$ decays.
It must be emphasised that due to the absence of loop contributions to the decay it is extremely clean theoretically~\cite{Brod:2013sga}.
This, and the abundance of different final states accessible, means that all parameters can be determined from data.
The relevant parameters are the weak phase $\gamma$, an associated strong (\CP conserving) phase difference between the $b \to c \bar{u}s$ and $b \to u \bar{c} s$ decay amplitudes, labelled $\delta_B$, and the ratio of their magnitudes, $r_B$. 
The small value of $r_B(B^- \to DK^-) \sim 10\,\%$ means that large event samples are necessary to obtain good constraints on $\gamma$, and only recently has the first $5\sigma$ observation of \CP violation in $B \to DK$ decays been achieved~\cite{LHCb-PAPER-2012-001}.
Larger values of $r_B$ are expected in $B^0 \to DK^{*0}$ and $\Bs \to \Dsmp\Kpm$ decays, but until now the samples available in these channels have not been been sufficient to give meaningful constraints on $\gamma$.
The available measurements use $B^{(*)-} \to D^{(*)}K^{(*)-}$ decays, with the latest combinations from each experiment giving (BaBar) $\gamma = (69\,^{+17}_{-16})^\circ$~\cite{Lees:2013zd}, (Belle) $\gamma = (68\,^{+15}_{-14})^\circ$~\cite{Trabelsi:2013uj} and (LHCb) $\gamma = (71\,^{+15}_{-16})^\circ$~\cite{LHCb-PAPER-2013-020}.
Significant progress in this area is anticipated from LHCb in the coming years.\footnote{
  Updates using more data have already started to appear from LHCb~\cite{LHCb-CONF-2013-004,LHCb-CONF-2013-006}.
}

\subsection{Mixing and \CP violation in the \Bs system}

A complete analysis of the time-dependent decay rates of neutral $B$ mesons must also take into account the lifetime difference between the eigenstates of the effective Hamiltonian, denoted by $\Delta \Gamma$.
This is particularly important in the $\Bs$ system, since the value of $\Delta \Gamma_s$ is non-negligible.
Neglecting $\CP$ violation in mixing, the relevant replacements for Eq.~(\ref{eq:SandC}) are~\cite{Dunietz:2000cr}
\begin{equation}
  \label{eq:SandCfull}
  \begin{array}{lcr}
    \multicolumn{2}{l}{
      \Gamma(\Bsb\,_{\rm phys} \to f(t)) = 
      {\cal N} 
      \frac{e^{-t/\tau(\Bs)}}{4\tau(\Bs)}
      \Big[ 
      \cosh(\frac{\Delta \Gamma t}{2}) +
    } & \hspace{40mm} \\
    \hspace{20mm} &
    \multicolumn{2}{r}{
      S_f \sin(\Delta m t) - C_f \cos(\Delta m t) +
      A^{\Delta \Gamma}_f \sinh(\frac{\Delta \Gamma t}{2})
      \Big],
    } \\
    \multicolumn{2}{l}{
      \Gamma(\Bs\,_{\rm phys} \to f(t)) =
      {\cal N} 
      \frac{e^{-t/\tau(\Bs)}}{4\tau(\Bs)}
      \Big[ 
      \cosh(\frac{\Delta \Gamma t}{2}) -
    } & \hspace{40mm} \\
    \hspace{20mm} & 
    \multicolumn{2}{r}{
      S_f \sin(\Delta m t) + C_f \cos(\Delta m t) +
      A^{\Delta \Gamma}_f \sinh(\frac{\Delta \Gamma t}{2})
      \Big]. 
    } \\
  \end{array}
\end{equation}
where ${\cal N}$ is a normalisation factor and 
\begin{equation}
  A^{\Delta \Gamma}_f = - \frac{2\, {\rm Re}(\lambda_f)}{1 + |\lambda_f|^2} \, .
\end{equation}
Note that, by definition,
\begin{equation}
  \left( S_f \right)^2 + \left( C_f \right)^2 + \left( A^{\Delta \Gamma}_f \right)^2 = 1 \, .
\end{equation}
Also $A^{\Delta \Gamma}_f$ is a \CP-conserving parameter, unlike $S_f$ and $C_f$ (since it appears with the same sign in equations for both $\Bsb$ and $\Bs$ states).
Nonetheless, it provides sensitivity to ${\rm arg}(\lambda_f)$, which means that interesting results can be obtained from untagged time-dependent analyses ({\it a.k.a.} effective lifetime measurements~\cite{Fleischer:2011cw}).

\begin{figure}[!htb]
\centering
\includegraphics[width=0.75\textwidth,bb=30 190 610 590,clip=true]{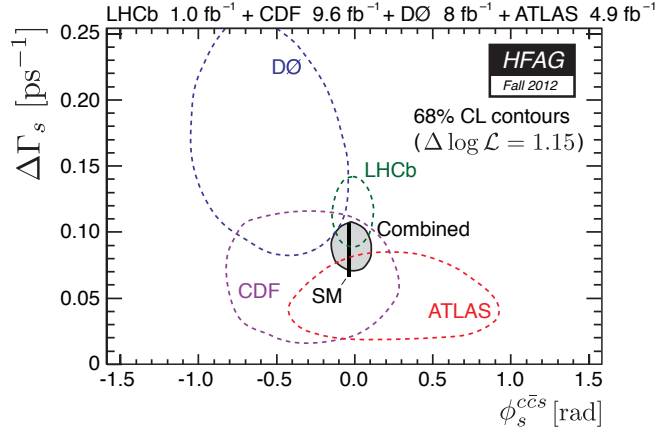}
\caption{
  Compilation of the latest results~\cite{Amhis:2012bh} in the $\phi_s$ {\it vs.} $\Delta \Gamma_s$ plane from LHCb~\cite{LHCb-CONF-2012-002},
  CDF~\cite{Aaltonen:2012ie}, D0~\cite{Abazov:2011ry} and ATLAS~\cite{Aad:2012kba}.
}
\label{fig:HFAG-phis}
\end{figure}

The formalism of Eq.~(\ref{eq:SandCfull}) is usually invoked for the determination of the \CP violation phase in \Bs oscillations, $\phi_s = -2\beta_s$, using $\Bs \to \jpsi\phi$ decays.
However, in that case things are complicated even further by the fact that the final state, containing two vector particles, is an admixture of \CP-even and \CP-odd which must be disentangled by angular analysis.\footnote{
  A somewhat more straightforward analysis can be done with the $\Bs \to \jpsi f_0(980)$ decay~\cite{LHCb-PAPER-2012-006}.
}  
Moreover, there is a potential contribution from S-wave $\Kp\Km$ pairs within the $\phi$ mass window used in the analysis.
However, all of these features can be turned to the benefit of the analysis, providing better sensitivity and allowing to resolve an ambiguity in the results~\cite{LHCb-PAPER-2011-028}.
A compilation of the latest results is shown in Fig.~\ref{fig:HFAG-phis}.\footnote{
    Very recent LHCb~\cite{LHCb-PAPER-2013-002} and ATLAS~\cite{ATLAS:2013nla} updates are not included.
}
Although great progress has been made over the last few years, the experimental precision does not yet challenge the theoretical uncertainty, and so further updates are of great interest.

\subsection{Mixing-induced \CP violation in hadronic $b \to s$ penguin decay modes}

As discussed in Sec.~\ref{subsec:dcpv}, decay modes mediated by penguin diagrams are potentially sensitive to NP effects, although it is a considerable challenge to disentangle QCD effects.
One useful approach is to study mixing-induced \CP violation effects in channels that are dominated by the penguin transition, so that little or no tree (or other) contribution is expected.
Such channels include $\Bz \to \phi\KS$, $\Bz\to\etapr\KS$, $\Bz \to \KS\KS\KS$, $\Bs \to \phi\phi$ and $\Bs \to \Kstarz\Kstarzb$.\footnote{
  The decay $\Bz\to\KS\piz$ is also of great interest since the tree contribution can be controlled using isospin relations to other $B\to K\pi$ decays.
}
For the $\Bz$ decays, the formalism is the same as given in Eq.~(\ref{eq:SandC}), and the parameters are expected in the SM to be given, to good approximations, by $C_f \approx 0$, $S_f \approx \sin(2\beta)$ (up to a sign, given by the \CP eigenvalue of the final state).
These channels have been studied extensively by BaBar and Belle: early results provided hints for discrepancies with the SM predictions, but the significance of the deviation diminshed with improved results~\cite{Lees:2012kxa,Nakahama:2010nj,Aubert:2008ad,Chen:2006nk,Lees:2011nf}.
For the $\Bs$ decays, the formalism is as given in Eq.~(\ref{eq:SandCfull}) (though with modifications due to the vector-vector final states), and the SM expectation is that \CP violation effects vanish, to a good approximation, since the very small phase in the $b \to s$ decay cancels that in the $\Bs$--$\Bsb$ oscillations.
First results have been reported by LHCb~\cite{LHCb-PAPER-2011-012,LHCb-PAPER-2012-004,LHCb-PAPER-2013-007}, and will reach a very interesting level of sensitivity as more data is accumulated.

\subsection{Charm mixing and \CP violation}

In the charm system the mixing parameters $x = \Delta m/\Gamma$ and $y = \Delta \Gamma/(2\Gamma)$ are both small, $x,y \ll 1$.
Therefore, a Taylor expansion can be performed on the generic expression of Eq.~(\ref{eq:SandCfull}) to give
\begin{equation}
  \label{eq:SandCcharm}
  \begin{array}{lcr}
    \Gamma(\Dzb\,_{\rm phys} \to f(t)) 
    & = &
    {\cal N}
    \frac{e^{-t/\tau(\Dz)}}{4\tau(\Dz)}
    \Big[ 
      1 - C_f + \left( S_f x + A^{\Delta \Gamma}_f y \right) \Gamma t
    \Big], \\
    \Gamma(\Dz\,_{\rm phys} \to f(t)) & = & 
    {\cal N}\frac{e^{-t/\tau(\Dz)}}{4\tau(\Dz)}
    \Big[ 
      1 + C_f - \left( S_f x - A^{\Delta \Gamma}_f y \right) \Gamma t
    \Big]. 
  \end{array}
\end{equation}

\begin{figure}[!htb]
  \centering
  \includegraphics[width=0.4\textwidth]{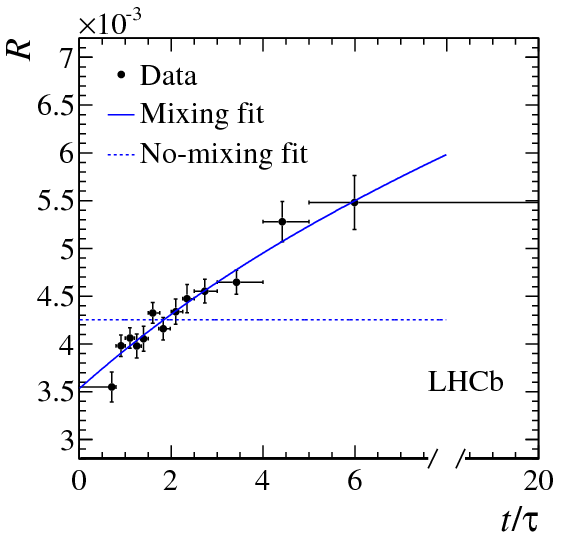}
  \includegraphics[width=0.4\textwidth]{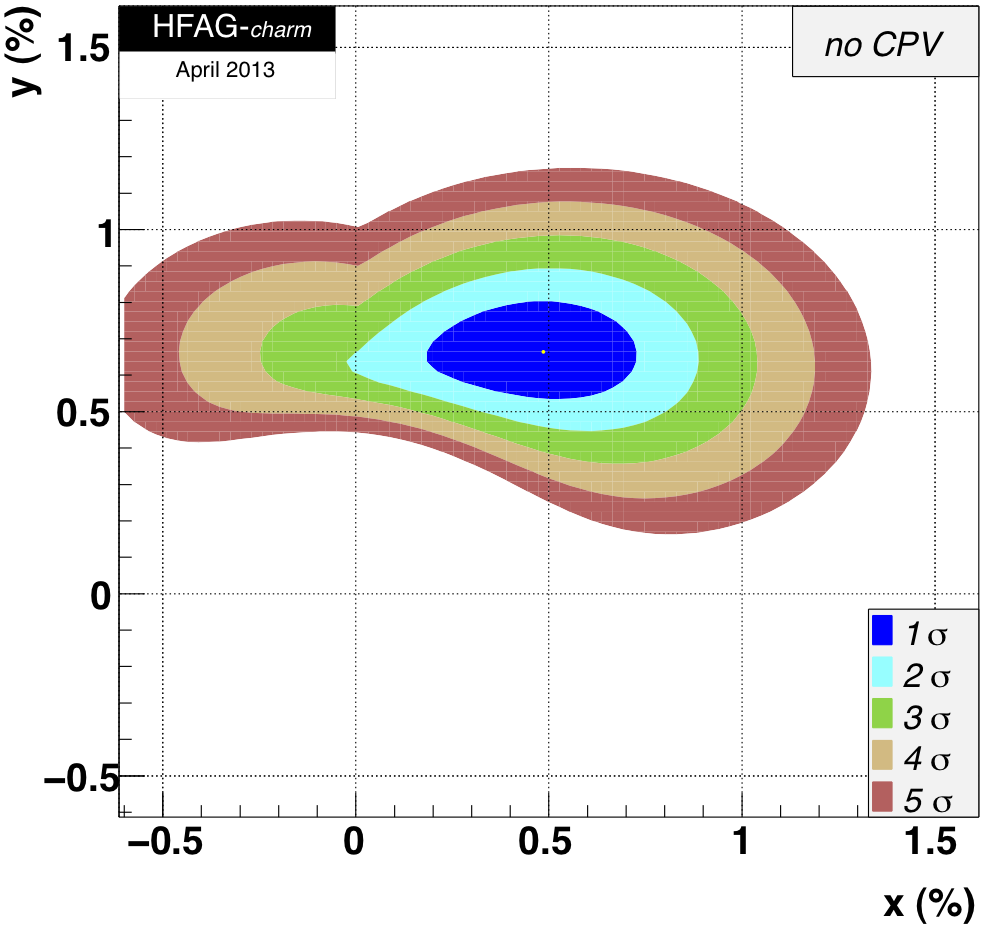}
  \caption{
    (Left) Decay-time evolution of the ratio, $R$, of $\Dz \to \Kp\pim$ to $\Dz \to \Km\pip$ yields (points) with the projection of the mixing allowed (solid line) and no-mixing (dashed line) fits overlaid, from Ref.~\cite{LHCb-PAPER-2012-038};
    (Right) World average constraints on the $x$ and $y$ parameters in the \Dz system~\cite{Amhis:2012bh}.
  }
\label{fig:charmMix}
\end{figure}

Hence an untagged analysis of $\Dz \to \Kp\Km$ can measure $A^{\Delta \Gamma}_f y$ (also known as $y_{\CP}$), while a tagged analysis can additionally probe $S_f x$.
Since the mixing parameters are small, the focus until now has been to establish definitively oscillation effects, but in the coming years the main objective will be to observe or limit \CP violation in the charm system, which is expected to be very small in the SM.
Note that in case the source of \Dz mesons is either from \Dstarp decays or semileptonic $b$-hadron decays, the flavour tagging is very effectively achieved from the charge of the associated pion or lepton, respectively.
Many other final states can be used to gain additional sensitivity to charm mixing and \CP violation parameters, a recent example being the observation of charm mixing at LHCb using $\Dz \to \Kp\pim$ decays~\cite{LHCb-PAPER-2012-038}.
The result of this analysis, and the world average constraints on the $x$ and $y$ parameters in the \Dz system,\footnote{
  Note that in Fig.~\ref{fig:charmMix}, the definition of the $x$ and $y$ parameters in the charm system is different from that in Sec.~\ref{subsec:RD} -- in this definition the \CP violating phase in \Dz oscillations is assumed to be small, and $x$ can be either positive or negative.
} 
are shown in Fig.~\ref{fig:charmMix}.

Direct \CP violation in the charm system can also be used to test the SM.
One interesting recent result has been the measurement of $\Delta A_{\CP}$, which is the difference between the direct \CP violation parameters of $\Dz \to \Kp\Km$ and $\Dz \to \pip\pim$ decays.
By measuring the difference, a cancellation of production and detection asymmetries can be exploited, while the physical \CP asymmetry may be enhanced.\footnote{
  The \CP asymmetries in $\Dz \to \Kp\Km$ and $\Dz \to \pip\pim$ decays are expected to be of opposite sign due to U-spin symmetry.
}
This method was first used by LHCb~\cite{LHCb-PAPER-2011-023} and then by CDF~\cite{Aaltonen:2012qw} and Belle~\cite{Ko:ICHEP}, all indicating a larger than expected effect.
This prompted a great deal of theoretical activity, summarised in Ref.~\cite{LHCb-PAPER-2012-031}, with the conclusion that a SM origin of the \CP violation, although unlikely, was not ruled out.
Many further studies were proposed to test both SM and NP hypotheses, and these remain of great interest and will be pursued.
However, the most recent results by LHCb~\cite{LHCb-PAPER-2013-003,LHCb-CONF-2013-003} suggest that the central value is smaller than previous thought, and therefore the SM explanation becomes harder to rule out.

\subsection{Photon polarisation in radiative \B decays}

The $b \to s\gamma$ transition is an archetypal flavour-changing neutral-current (FCNC) transition, and has been considered a sensitive probe for NP since the first measurements of its rate~\cite{Ammar:1993sh,Alam:1994aw}.
The latest results for the inclusive branching fraction~\cite{Amhis:2012bh} are consistent with the SM prediction~\cite{Misiak:2006zs}
\begin{eqnarray}
  {\cal B}\left( B \to X_s \gamma \right)^{\rm exp}_{E_\gamma > 1.7 \gev} & = & ( 3.43 \pm 0.21 \pm 0.07 ) \times 10^{-4} \, , \\
  {\cal B}\left( B \to X_s \gamma \right)^{\rm th}_{E_\gamma > 1.7 \gev} & = & ( 3.15 \pm 0.23 ) \times 10^{-4} \, .
\end{eqnarray}
However, additional observables, such as \CP and isospin asymmetries provide complementary sensitivity and still have experimental uncertainties much larger than those of the theoretical predictions of their values in the SM.

One particularly interesting observable is the polarisation of the emitted photon in $b \to s \gamma$ decays, since the $V-A$ structure of the SM weak interaction results in a high degree of polarisation, that is not necessarily reproduced in extended models.
Until now, the most promising approach to probe the polarisation has been from time-dependent asymmetry measurements of $\Bz \to \KS\piz\gamma$~\cite{Atwood:1997zr,Atwood:2004jj} but the most recent measurements~\cite{Aubert:2008gy,Ushiroda:2006fi} still have large uncertainties.
LHCb can use several different methods to study photon polarisation in $b \to s\gamma$ transitions, such as measuring the effective lifetime in $\Bs \to \phi\gamma$ decays~\cite{Muheim:2008vu}.
Although all such measurements are highly challenging, the observed yields in $\Bs \to \phi\gamma$~\cite{LHCb-PAPER-2012-019} and other related channels such as $\Bz \to \Kstarz\epem$~\cite{LHCb-PAPER-2013-005} suggest there are good prospects for significant progress in the coming years.

\subsection{Angular observables in $\Bz \to \Kstarz \mu^+\mu^-$ decays}
\label{sec:kstarmumu}

The $b \to s l^+l^-$ FCNC transitions offer similar, but complementary, sensitivity to NP as $b \to s\gamma$, but are experimentally more convenient to study, in particular when the lepton pair is muonic, \ie\ $l^+l^- = \mu^+\mu^-$.
The multi-body final state provides access to a wide range of kinematic observables, several of which have clean theoretical predictions (especially at low values of the dilepton invariant mass squared, $q^2$).
This makes these decays a superb laboratory for NP tests.

The theoretical framework for these (and other) processes is the operator product expansion.
This is an effective theory, applicable for $b$ physics, which describes the weak interactions of the SM by integrating out the heavier ($W$, $Z$, $t$) fields.
As such it can be considered a modern version of the Fermi theory of beta decay.
Conceptually, it can be expressed as
\begin{equation}
  {\cal L}_{(\rm full \ EW \times QCD)} \longrightarrow 
  {\cal L}_{\rm eff} = {\cal L}_{\rm QED \times QCD} {\small \left(\begin{array}{c}{\rm quarks \neq t}\\{\rm leptons}\end{array}\right)} + \Sigma_n C_n {\cal O}_n \, ,
\end{equation}
where ${\cal O}_n$ represent the local interaction terms, and $C_n$ are coupling constants that are referred to as Wilson coefficients.\footnote{
  As written here the $C_n$ include the Fermi coupling and the CKM matrix elements, but usually these terms are factored out.
}
The Wilson coefficients encode information on the weak scale, and are calculable and known in the SM (at least to leading order).
Moreover, they are affected by NP, so comparing the measured values with their expectations provides tests of the SM.
A more detailed description of the operator product expansion can be found in, \eg\ Ref.~\cite{Buras:2005xt}.

For the purposes of discussing $b \to s l^+l^-$ decays, the Wilson coefficients of interest are $C_7$ (which also affects $b \to s\gamma$), $C_9$ and $C_{10}$.
The differential decay distribution, for the inclusive process, can be written~\cite{Lee:2006gs}
\begin{equation}
  \label{eq:btosmumu}
  \frac{d^2\Gamma}{dq^2\,d\cos\theta_l} = \frac{3}{8}\left[
    (1+\cos^2\theta_l)H_T(q^2) + 2 \cos\theta_l H_A(q^2) + 2(1-\cos^2\theta_l)H_L(q^2)
  \right]
\end{equation}
where $\theta_l$ is the angle between the momentum vectors of the positively charged lepton and the opposite of the decaying $b$ hadron in the dilepton rest frame.\footnote{
  The full decay distribution for $\Bz\to \Kstarz \mu^+\mu^-$ and other $B \to Vl^+l^-$ ($V = \rho, \omega, \Kstar, \phi$) decays includes two other angles: the decay angle of the vector meson (usually denoted $\theta_{V}$) and the angle between the two decay planes (usually denoted $\phi$).
}
The coefficients are given by 
\begin{eqnarray}
  \label{eq:HTAL}
  H_T(q^2) & \propto & 2q^2\left[\left( C_9 + 2C_7 \frac{m_b^2}{q^2} \right)^2 + C_{10}^2 \right] \nonumber \\
  H_A(q^2) & \propto & -4q^2C_{10}\left( C_9 + 2C_7 \frac{m_b^2}{q^2} \right) \\
  H_L(q^2) & \propto & \left[ \left( C_9 + 2C_7\right)^2 + C_{10}^2 \right] \, . \nonumber
\end{eqnarray}
Note that the term involving $H_A$ depends linearly on $\cos \theta_l$ and hence gives rise to a $q^2$-dependent forward backward asymmetry, $A_{\rm FB}$.
The shape of $A_{\rm FB}$, in particular the value of $q^2$ at which it crosses zero, can be predicted with low uncertainty in the SM.
The expressions for exclusive processes, such as $\Bz\to\Kstarz\mu^+\mu^-$, are conceptually similar to those of Eqs.~(\ref{eq:btosmumu}) and~(\ref{eq:HTAL}), but are more complicated as they also involve hadronic form factors.
On the other hand, exclusive channnels also provide additional observables that can be studied (such as the longitudinal polarisation of the $\Kstarz$ meson, $F_{\rm L}$), some of which can be precisely predicted in the SM, and are sensitive to NP contributions.

\begin{figure}[!htb]
  \centering
  \includegraphics[width=0.4\textwidth]{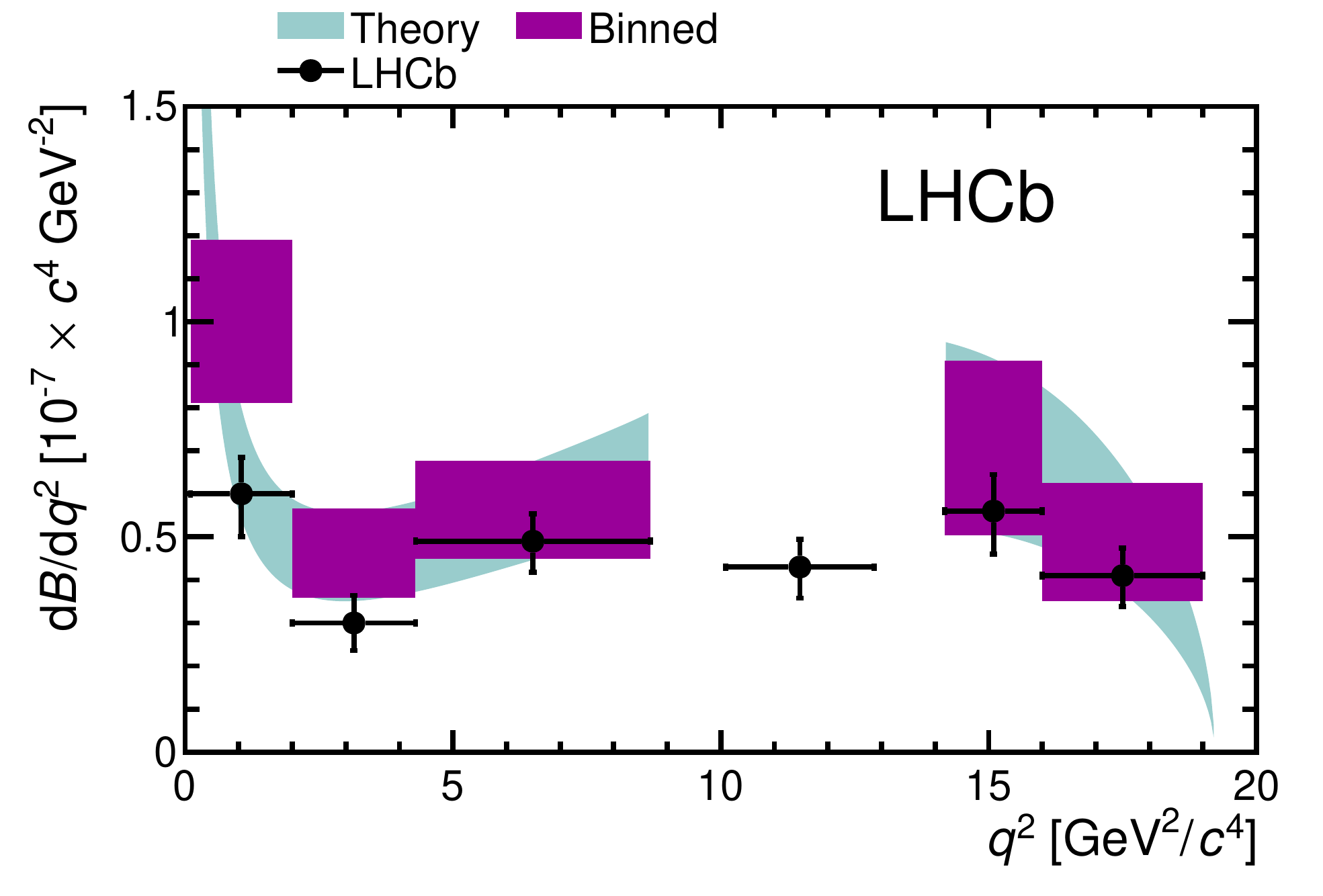}
  \includegraphics[width=0.4\textwidth]{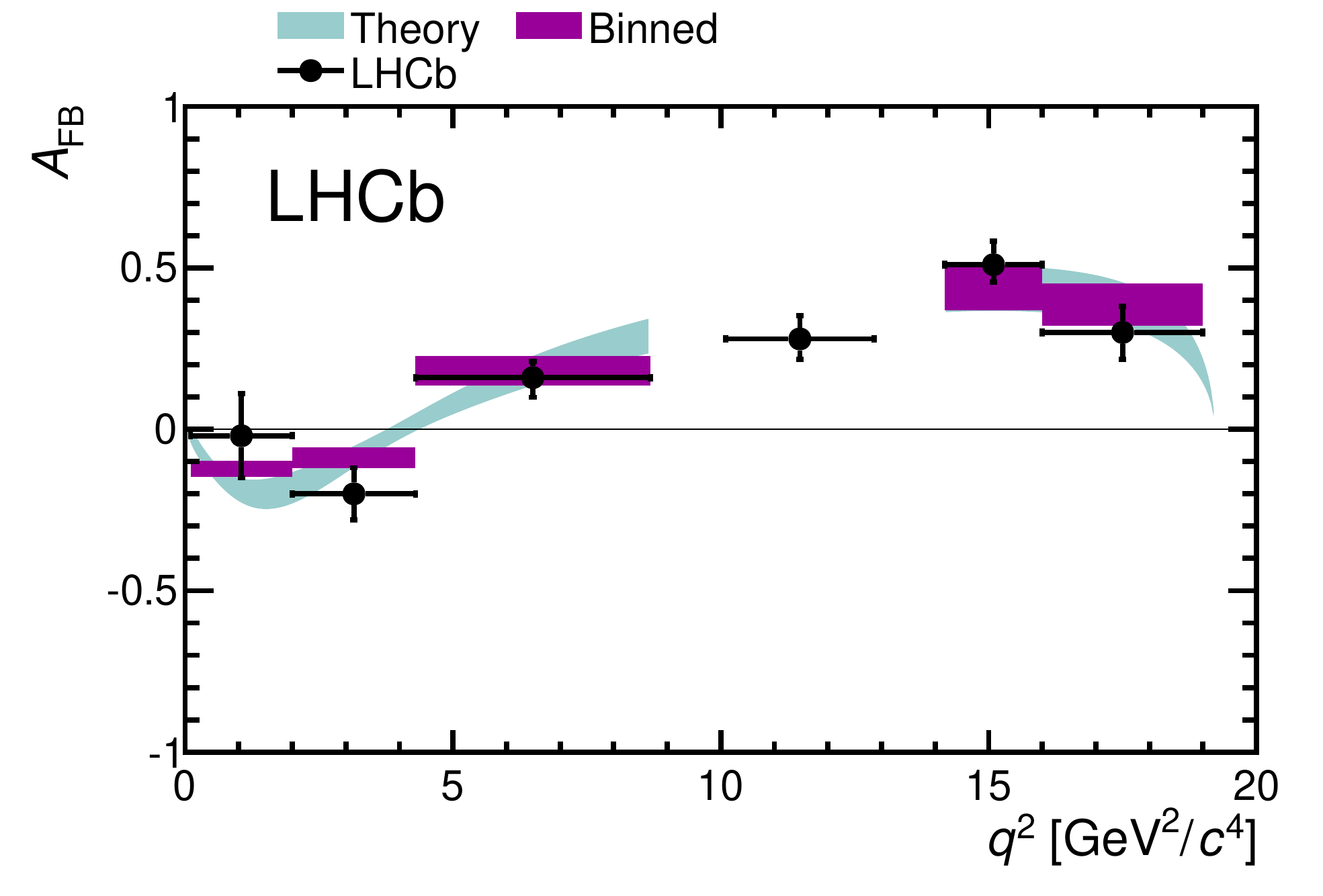}
  \caption{
    (Left) Differential branching fraction and (right) $A_{\rm FB}$ of $\Bz\to\Kstarz\mu^+\mu^-$ decays in bins of $q^2$ as measured by LHCb~\cite{LHCb-PAPER-2013-019}.
  }
  \label{fig:Kstarmumu}
\end{figure}

The decay rates and angular distributions of $\Bz\to\Kstarz\mu^+\mu^-$ decays have been studied by many experiments, with the most precise results to date, from LHCb~\cite{LHCb-PAPER-2013-019}, shown in Fig.~\ref{fig:Kstarmumu}.
This analysis provides the first measurement of the $A_{\rm FB}$ zero crossing point, $q_{0}^{2} = 4.9 \pm 0.9 \,{\rm GeV}^{2}/c^{4}$, consistent with the SM prediction.
Significant progress, including improved measurements of other NP-sensitive angular observables, can be expected in the coming years.

\subsection{The very rare decay $\Bs \to \mu^+\mu^-$}

The ``killer app.'' for flavour physics as a tool to probe for (and potentially discover) NP is the very rare decay $\Bs \to \mu^+\mu^-$.
The branching fraction is highly suppressed in the SM due to a combination of three factors, none of which are necessarily reproduced in extended models: (i) the absence of tree-level FCNC transitions; (ii) the $V-A$ structure of the weak interaction that results in helicity suppression of purely leptonic decays of flavoured pseudoscalar mesons; (iii) the hierarchy of the CKM matrix elements.
In particular, in the minimally supersymmetric extension of the SM, the presence of a pseudoscalar Higgs particle alleviates the helicity suppression and enhances the branching fraction by a factor proportional to $\tan^6\beta/M_{A_0}^4$, where $\tan\beta$ is the ratio of Higgs' vacuum expectation values, and $M_{A_0}$ is the pseudoscalar Higgs mass.
Therefore, in the region of phase-space where $\tan\beta$ is not too small, and $M_{A_0}$ is not too large, the decay rate can be significantly enhanced above its SM expectation~\cite{Buras:2012ru},\footnote{
  Note that, due to the non-zero value of the decay width difference in the \Bs system, this value needs to be corrected upwards by $\sim 9\%$ to obtain the experimentally measured (\ie, decay time integrated) quantity~\cite{DeBruyn:2012wk}.
}
\begin{equation}
  {\cal B}\left(\Bs \to \mu^+\mu^-\right)^{\rm SM} = \left( 3.2 \pm 0.3 \right) \times 10^{-9} \, .
\end{equation}

\begin{figure}[!htb]
  \centering
  \includegraphics[width=0.6\textwidth]{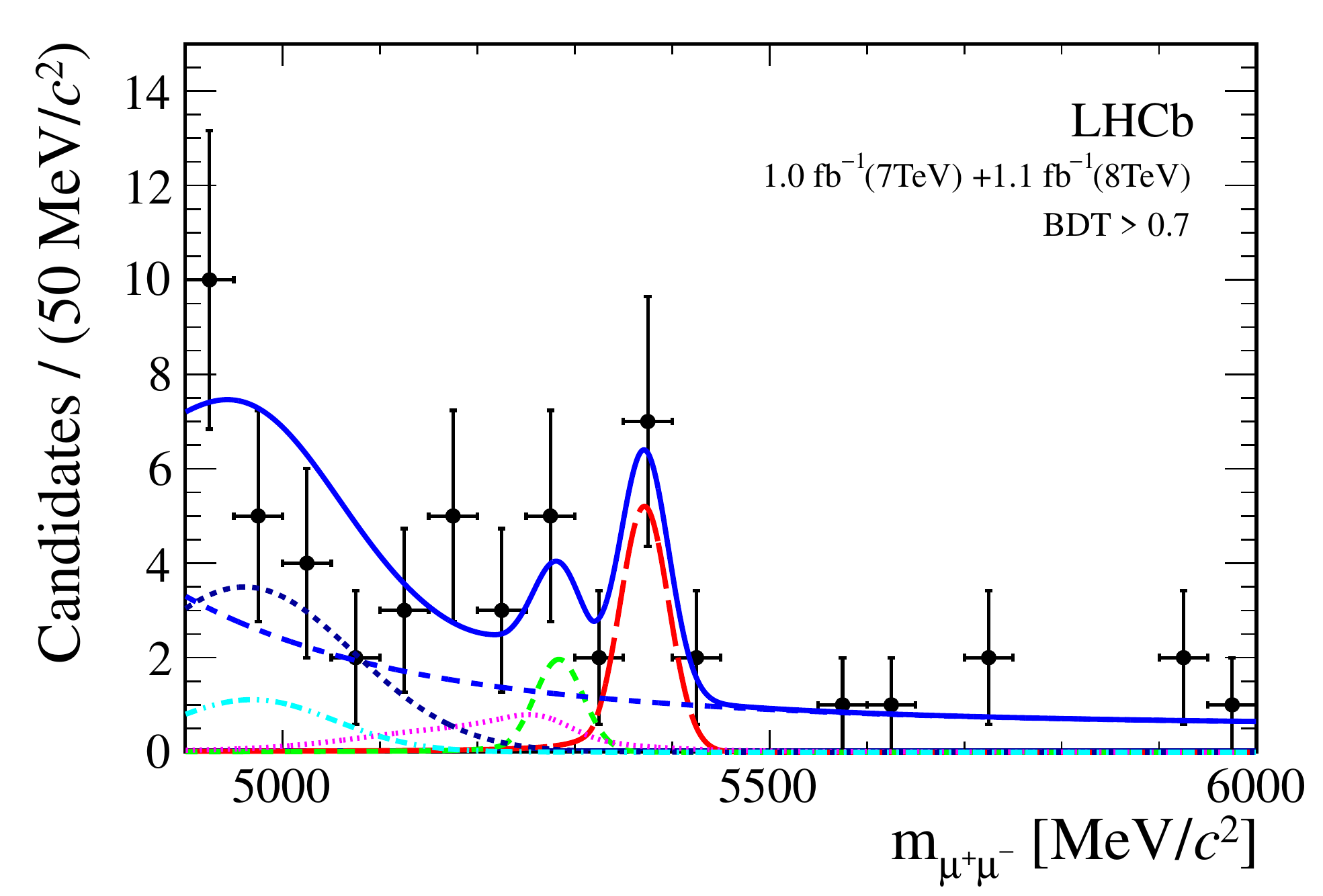}
  \caption{
    Invariant mass distribution of selected $\Bs \to \mu^+\mu^-$ candidates, with fit result overlaid~\cite{LHCb-PAPER-2012-043}.
  }
  \label{fig:Bsmumu}
\end{figure}

Due to the very clean signature of this decay, it has been studied by essentially all high-energy hadron collider experiments.
The crucial components to obtain good sensitivity are high luminosity, a large $B$ production cross-section within the acceptance, and good vertex and mass resolution to reject the background.
Although ATLAS~\cite{Aad:2012pn} and CMS~\cite{Chatrchyan:2012rga} have collected more luminosity, at present the strengths of the LHCb detector have allowed it to obtain the most precise results for this mode.
Following a series of increasingly restrictive upper limits~\cite{LHCb-PAPER-2011-004,LHCb-PAPER-2011-025,LHCb-PAPER-2012-007}, LHCb has recently obtained the first evidence, with $3.5\sigma$ significance, for the decay~\cite{LHCb-PAPER-2012-043}, as shown in Fig.~\ref{fig:Bsmumu}.
The branching fraction is measured to be
\begin{equation}
  {\cal B}\left(\Bs \to \mu^+\mu^-\right) = \left( 3.2 \,^{+1.4}_{-1.2} ({\rm stat})\,^{+0.5}_{-0.3} ({\rm syst})\right) \times 10^{-9} \, ,
\end{equation}
in agreement with the SM prediction.

Further updates of this measurement are keenly anticipated, and are likely to appear at regular intervals throughout the lifetime of the LHC.
It is worth noting that even in case the $\Bs \to \mu^+\mu^-$ branching fraction remains consistent with the SM, the decay provides an additional handle on NP through its effective lifetime~\cite{Buras:2013uqa}.
Moreover, it will be important to study also the even more suppressed $\Bd \to \mu^+\mu^-$ decay, since the ratio of the $\Bd$ and $\Bs$ branching fractions is a benchmark test of MFV.

\section{Future flavour physics experiments}

\begin{figure}[!htb]
  \centering
  \includegraphics[width=0.36\textwidth]{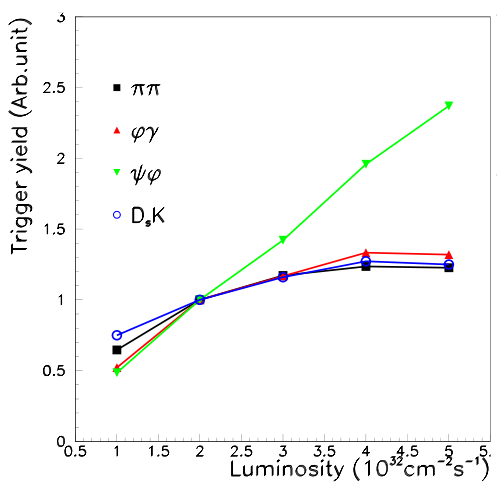}
  \hspace{3mm}
  \includegraphics[width=0.55\textwidth]{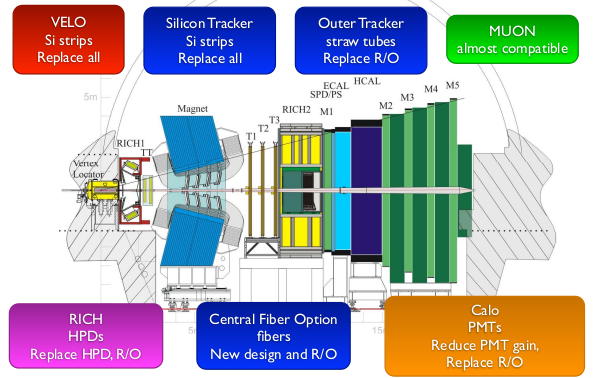}
  \caption{
    (Left) Scaling of yields with instantaneous in certain decay channels at LHCb~\cite{CERN-LHCC-2012-007}, showing the limitation caused by the $1 \mhz$ readout.
    Note that during 2012 LHCb operated at an instantaneous luminosity of $4 \times 10^{32} \cm^{-2} \sec^{-1}$.
    (Right) Illustration of the key components of the LHCb subdetector upgrades.
  }
  \label{fig:hans}
\end{figure}

As stressed in the previous sections, the first results from the LHC have already provided dramatic advances in flavour physics, and significant further progress is anticipated in the coming years.
However, the instantaneous luminosity of LHCb is limited due to the fact that its subdetectors are read out at $1\mhz$.
As shown in Fig.~\ref{fig:hans} (left), increasing the luminosity requires tightening of the hardware trigger thresholds in order not to exceed this limit.
This then results in lower efficiencies, especially for decay channels triggered by the calorimeter (\ie, channels without muons in the final state), so that there is no net gain in yield.
Therefore, after several years of operation at the optimal instantaneous luminosity at $\sqrt{s} = 13 \ {\rm or} \ 14 \tev$,\footnote{
  Note that heavy flavour cross-sections increase with $\sqrt{s}$, so a significant boost in yields is expected when moving to higher energies.
} the time required to double the accumulated statistics becomes excessively long.

As should be clear from the discussions above, it remains of great importance to pursue a wide range of flavour physics measurements and improve their precision to the level of the theoretical uncertainty, and therefore it is of clear interest to get past the $1 \mhz$ readout limitation.
The concept of the LHCb upgrade~\cite{CERN-LHCC-2011-001,CERN-LHCC-2012-007} is to read out the full detector at $40\mhz$ (which corresponds to the maximum bunch crossing rate, with $25 \ns$ spacing) and implement the trigger fully in software.
This will allow the experiment to run at higher luminosities, up to $1 \ {\rm or} \ 2 \times 10^{33} \cm^{-2} \sec^{-1}$, and will also significantly improve the efficiency for modes currently triggered by calorimeter signals at the hardware level.
The accumulated samples in most key modes will increase by around two orders of magnitude compared to what was collected in 2011.
Moreover, with a flexible trigger scheme, the capability to search for other signatures of NP will be enhanced, so that the upgraded experiment can be considered a general purpose detector in the forward region.
The LHCb upgrade is planned to occur during the second long shutdown of the LHC, in 2018.
Since its target luminosity is still below that which can be delivered by the LHC, it does not depend (though it is consistent with) the HL-LHC machine upgrade.

There are several other flavour physics experiments that will be coming online on a similar same timescale.
The KEKB accelerator and Belle experiment are being upgraded~\cite{Abe:2010sj}, in order to allow luminosities almost two orders of magnitude larger than has previously been achieved.
Compared to the LHCb upgrade, the $\epem$ environment is advantageous for decay modes with missing energy and for inclusive measurements.
Some of the key channels for Belle2 are lepton flavour violating decays of $\tau$ leptons, mixing-induced \CP asymmetries in decays such as $\Bz \to \phi\KS$ and $\Bz \to \etapr \KS$, and the leptonic decay $\Bp \to \taup\nu$ (which can be considered a counterpart of $\Bs\to\mu^+\mu^-$, and is sensitive to the exchange of charged Higgs particles)~\cite{Browder:2008em,Aushev:2010bq}.

In addition, the NA62~\cite{NA62-10-07} and K0T0~\cite{Yamanaka:2012yma} experiments will search for the $\Kp \to \pip\neu\neub$ and $\KL\to\piz\neu\neub$ decays, respectively.
Long considered the ``holy grail'' of kaon physics these decays are highly suppressed in the SM and have clean theoretical predictions.
The new generation of experiments should be able to observe these channels for the first time, if they occur at around the SM rate.

\section{Conclusion}

Flavour physics continues to present many mysteries, and these demand continued experimental and theoretical investigation.
Heavy flavour physics is complementary to other sectors of the global particle physics programme such as the high-\pt experiments at the LHC, and neutrino oscillation and low energy precision experiments.
The prospects are good for significant progress in the coming few years and, with upgraded experiments planned to come online in the second half of this decade, beyond.

\begin{acknowledgement}
  These lectures were delivered, in variously modified forms, at the Hadron Collider Physics Summer School 2010 in Fermilab, USA, the 2012 Spring School ``Bruno Touschek'' of the Frascati National Laboratories in Frascati, Italy and the $69^{\rm th}$ Scottish Universities Summer School in Physics in St. Andrews, UK, 2012.
  The author is grateful to the organisers of these meetings for invitations and support, and to the participants for stimulating discussions and questions.
  This work is supported by the 
  Science and Technology Facilities Council (United Kingdom), CERN, 
  and by the European Research Council under FP7.
\end{acknowledgement}

\bibliographystyle{spphys}
\bibliography{gershon-SUSSP,LHCb-PAPER}

\end{document}

%% file: lhcb-symbols-def.tex
\def\ux85 {\mbox{UX85}\xspace}

\ifthenelse{\boolean{uprightparticles}}%
{

 \def\Peta        {\ensuremath{\upeta}\xspace}

 \def\Pmu         {\ensuremath{\upmu}\xspace}                 
 \def\Pnu         {\ensuremath{\upnu}\xspace}                 
                  
 \def\Ppi         {\ensuremath{\uppi}\xspace}

 \def\Ptau        {\ensuremath{\uptau}\xspace}

 \def\Ppsi        {\ensuremath{\uppsi}\xspace}

 \def\PDelta      {\ensuremath{\Delta}\xspace}                 
 \def\PXi      {\ensuremath{\Xi}\xspace}                 
 \def\PLambda      {\ensuremath{\Lambda}\xspace}                 
 \def\PSigma      {\ensuremath{\Sigma}\xspace}                 
 \def\POmega      {\ensuremath{\Omega}\xspace}                 
 \def\PUpsilon      {\ensuremath{\Upsilon}\xspace}                 
 
 \def\PB      {\ensuremath{\mathrm{B}}\xspace}                 
                  
 \def\PD      {\ensuremath{\mathrm{D}}\xspace}

 \def\PJ      {\ensuremath{\mathrm{J}}\xspace}                 
 \def\PK      {\ensuremath{\mathrm{K}}\xspace}

 \def\Pb      {\ensuremath{\mathrm{b}}\xspace}

 \def\Pe      {\ensuremath{\mathrm{e}}\xspace}

 \def\Pi      {\ensuremath{\mathrm{i}}\xspace}

 \def\Ps      {\ensuremath{\mathrm{s}}\xspace}

}
{

 \def\Peta        {\ensuremath{\eta}\xspace}

 \def\Pmu         {\ensuremath{\mu}\xspace}                 
 \def\Pnu         {\ensuremath{\nu}\xspace}                 
                  
 \def\Ppi         {\ensuremath{\pi}\xspace}

 \def\Ptau        {\ensuremath{\tau}\xspace}

 \def\Ppsi        {\ensuremath{\psi}\xspace}                 
                  
 \mathchardef\PDelta="7101
 \mathchardef\PXi="7104
 \mathchardef\PLambda="7103
 \mathchardef\PSigma="7106
 \mathchardef\POmega="710A
 \mathchardef\PUpsilon="7107
                  
 \def\PB      {\ensuremath{B}\xspace}                 
                  
 \def\PD      {\ensuremath{D}\xspace}

 \def\PJ      {\ensuremath{J}\xspace}                 
 \def\PK      {\ensuremath{K}\xspace}

 \def\Pb      {\ensuremath{b}\xspace}

 \def\Pe      {\ensuremath{e}\xspace}

 \def\Pi      {\ensuremath{i}\xspace}

 \def\Ps      {\ensuremath{s}\xspace}

}

\def\epem       {\ensuremath{\Pe^+\Pe^-}\xspace}

\def\mumu       {\ensuremath{\Pmu^+\Pmu^-}\xspace}

\def\taup       {\ensuremath{\Ptau^+}\xspace}

\def\neu        {\ensuremath{\Pnu}\xspace}
\def\neub       {\ensuremath{\overline{\Pnu}}\xspace}

\def\squark    {\ensuremath{\Ps}\xspace}

\def\bquark    {\ensuremath{\Pb}\xspace}

\def\pion  {\ensuremath{\Ppi}\xspace}
\def\piz   {\ensuremath{\pion^0}\xspace}

\def\pip   {\ensuremath{\pion^+}\xspace}
\def\pim   {\ensuremath{\pion^-}\xspace}

\def\kaon  {\ensuremath{\PK}\xspace}
  \def\Kbar  {\kern 0.2em\overline{\kern -0.2em \PK}{}\xspace}

\def\Kz    {\ensuremath{\kaon^0}\xspace}
\def\Kzb   {\ensuremath{\Kbar^0}\xspace}
\def\KzKzb {\ensuremath{\Kz \kern -0.16em \Kzb}\xspace}
\def\Kp    {\ensuremath{\kaon^+}\xspace}
\def\Km    {\ensuremath{\kaon^-}\xspace}
\def\Kpm   {\ensuremath{\kaon^\pm}\xspace}

\def\KpKm  {\ensuremath{\Kp \kern -0.16em \Km}\xspace}
\def\KS    {\ensuremath{\kaon^0_{\rm\scriptscriptstyle S}}\xspace} 
\def\KL    {\ensuremath{\kaon^0_{\rm\scriptscriptstyle L}}\xspace} 
\def\Kstarz  {\ensuremath{\kaon^{*0}}\xspace}
\def\Kstarzb {\ensuremath{\Kbar^{*0}}\xspace}
\def\Kstar   {\ensuremath{\kaon^*}\xspace}

\newcommand{\etapr}{\ensuremath{\Peta^{\prime}}\xspace}

  \def\Dbar    {\kern 0.2em\overline{\kern -0.2em \PD}{}\xspace}
\def\D       {\ensuremath{\PD}\xspace}

\def\Dz      {\ensuremath{\D^0}\xspace}
\def\Dzb     {\ensuremath{\Dbar^0}\xspace}
\def\DzDzb   {\ensuremath{\Dz {\kern -0.16em \Dzb}}\xspace}
\def\Dp      {\ensuremath{\D^+}\xspace}
\def\Dm      {\ensuremath{\D^-}\xspace}

\def\DpDm    {\ensuremath{\Dp {\kern -0.16em \Dm}}\xspace}

\def\Dstarp  {\ensuremath{\D^{*+}}\xspace}

\def\Dsmp    {\ensuremath{\D^{\mp}_\squark}\xspace}

\def\B       {\ensuremath{\PB}\xspace}
  \def\Bbar    {\kern 0.18em\overline{\kern -0.18em \PB}{}\xspace}

\def\Bz      {\ensuremath{\B^0}\xspace}
\def\Bzb     {\ensuremath{\Bbar^0}\xspace}
\def\Bu      {\ensuremath{\B^+}\xspace}
\def\Bub     {\ensuremath{\B^-}\xspace}
\def\Bp      {\ensuremath{\Bu}\xspace}
\def\Bm      {\ensuremath{\Bub}\xspace}
\def\Bpm     {\ensuremath{\B^\pm}\xspace}

\def\Bd      {\ensuremath{\B^0}\xspace}
\def\Bs      {\ensuremath{\B^0_\squark}\xspace}

\def\Bsb     {\ensuremath{\Bbar^0_\squark}\xspace}

\def\jpsi     {\ensuremath{{\PJ\mskip -3mu/\mskip -2mu\Ppsi\mskip 2mu}}\xspace}

  \def\Y#1S{\ensuremath{\PUpsilon{(#1S)}}\xspace}

\def\Lbar {\ensuremath{\kern 0.1em\overline{\kern -0.1em\PLambda}}\xspace}

\def\Lb      {\ensuremath{\PLambda^0_\bquark}\xspace}

\def\to                 {\ensuremath{\rightarrow}\xspace}

\def\CP                {\ensuremath{C\!P}\xspace}
\def\CPT               {\ensuremath{C\!PT}\xspace}

\def\rhobar {\ensuremath{\overline \rho}\xspace}
\def\etabar {\ensuremath{\overline \eta}\xspace}

\def\AT#1     {\ensuremath{A_{\mathrm{T}}^{#1}}\xspace}           

\def\C#1      {\ensuremath{\mathcal{C}_{#1}}\xspace}                       
\def\Cp#1     {\ensuremath{\mathcal{C}_{#1}^{'}}\xspace}                    
\def\Ceff#1   {\ensuremath{\mathcal{C}_{#1}^{\mathrm{(eff)}}}\xspace}        
\def\Cpeff#1  {\ensuremath{\mathcal{C}_{#1}^{'\mathrm{(eff)}}}\xspace}       
\def\Ope#1    {\ensuremath{\mathcal{O}_{#1}}\xspace}                       
\def\Opep#1   {\ensuremath{\mathcal{O}_{#1}^{'}}\xspace}                    

\newcommand{\tev}{\ensuremath{\mathrm{\,Te\kern -0.1em V}}\xspace}
\newcommand{\gev}{\ensuremath{\mathrm{\,Ge\kern -0.1em V}}\xspace}
\newcommand{\mev}{\ensuremath{\mathrm{\,Me\kern -0.1em V}}\xspace}
\newcommand{\kev}{\ensuremath{\mathrm{\,ke\kern -0.1em V}}\xspace}
\newcommand{\ev}{\ensuremath{\mathrm{\,e\kern -0.1em V}}\xspace}
\newcommand{\gevc}{\ensuremath{{\mathrm{\,Ge\kern -0.1em V\!/}c}}\xspace}
\newcommand{\mevc}{\ensuremath{{\mathrm{\,Me\kern -0.1em V\!/}c}}\xspace}
\newcommand{\gevcc}{\ensuremath{{\mathrm{\,Ge\kern -0.1em V\!/}c^2}}\xspace}
\newcommand{\gevgevcccc}{\ensuremath{{\mathrm{\,Ge\kern -0.1em V^2\!/}c^4}}\xspace}
\newcommand{\mevcc}{\ensuremath{{\mathrm{\,Me\kern -0.1em V\!/}c^2}}\xspace}

\def\cm   {\ensuremath{\rm \,cm}\xspace}

\def\mub{\ensuremath{\rm \,\upmu b}\xspace}

\def\nb {\ensuremath{\rm \,nb}\xspace}

\def\invfb   {\ensuremath{\mbox{\,fb}^{-1}}\xspace}

\def\invab   {\ensuremath{\mbox{\,ab}^{-1}}\xspace}

\def\sec  {\ensuremath{\rm {\,s}}\xspace}

\def\ns   {\ensuremath{{\rm \,ns}}\xspace}

\def\mhz  {\ensuremath{{\rm \,MHz}}\xspace}
\def\khz  {\ensuremath{{\rm \,kHz}}\xspace}
\def\hz   {\ensuremath{{\rm \,Hz}}\xspace}

\def\gsim{{~\raise.15em\hbox{$>$}\kern-.85em
          \lower.35em\hbox{$\sim$}~}\xspace}
\def\lsim{{~\raise.15em\hbox{$<$}\kern-.85em
          \lower.35em\hbox{$\sim$}~}\xspace}

\def\pt         {\mbox{$p_{\rm T}$}\xspace}

\def\tell1  {TELL1\xspace}
\def\ukl1   {UKL1\xspace}

\newcommand{\eg}{\mbox{\itshape e.g.}\xspace}
\newcommand{\ie}{\mbox{\itshape i.e.}}

\newcommand{\etc}{\mbox{\itshape etc.}\xspace}